%% file: apssamp.tex
\documentclass[letterpaper,aps,pra,tightenlines,preprintnumbers,nofootinbib,showkeys,superscriptaddress,longbibliography,notitlepage,twocolumn]{revtex4-1}
\pdfoutput=1

\usepackage[dvipsnames]{xcolor}
\usepackage{graphicx,epic,eepic,epsfig}
\usepackage[export]{adjustbox}
\usepackage{tikz,contour} 
\usepackage[compat=1.1.0]{tikz-feynman}
\usetikzlibrary{arrows.meta,calc,quantikz2}
\usetikzlibrary{shapes,arrows,positioning,automata,backgrounds,er,patterns}
\usepackage{fullpage}
\usepackage{relsize}
\usepackage[left=1.6cm,right=1.6cm,top=2cm,bottom=2cm]{geometry}

\usepackage{XCharter}
\usepackage{mathptmx}
\usepackage[T1]{fontenc}
\usepackage{pifont}
\usepackage{amsfonts,amsmath,amssymb,amsbsy,bm,mathtools,slashed,latexsym,esvect}
\usepackage{float,multirow,array,cancel,dcolumn}
\usepackage{enumitem}
\usepackage{hyperref}
\hypersetup{colorlinks=true,citecolor=red,linkcolor=NavyBlue!70!blue,urlcolor=Brown}
\usepackage[caption=false,labelformat=simple]{subfig}
\usepackage{lipsum}

\usepackage{comment}
\usepackage{natbib}
\usepackage{braket}

\usepackage{booktabs}
\usepackage{dsfont}
\newcommand{\momentum}[1]{\boldsymbol{#1}}

\newcommand{\fermion}[2]{\boldsymbol{#1},#2}
\providecommand{\ket}[1]{\vert #1 \rangle}
\providecommand{\bra}[1]{\langle #1 \vert}
\providecommand{\braket}[2]{\langle #1 \vert #2 \rangle}
\newcommand{\ketbra}[2]{\vert #1 \rangle \! \langle #2 \vert}
\providecommand{\proj}[1]{\ketbra{#1}{#1}}
\newcommand{\Id}{\mathds{1}}

\begin{document}

\title{\textbf{Three-qubit entanglement in QED scattering with an entangled spectator} 
} 

\author{Debanil Chowdhury}
\email{debanil.chowdhury@research.iiit.ac.in}
\affiliation{Center for Quantum Science and Technology, \\International Institute of Information Technology, Hyderabad 500~032, India}
\affiliation{Center for Security, Theory and Algorithmic Research, 
International Institute of Information Technology, Hyderabad 500~032, India}

\author{Maaz Khan}
\email{maaz.khan@research.iiit.ac.in}
\affiliation{Center for Computational Natural Sciences and Bioinformatics, International Institute of Information Technology, Hyderabad 500~032, India}

\author{Indranil Chakrabarty}
\email{indranil.chakrabarty@iiit.ac.in}
\affiliation{Center for Quantum Science and Technology, \\International Institute of Information Technology, Hyderabad 500~032, India}
\affiliation{Center for Security, Theory and Algorithmic Research, 
International Institute of Information Technology, Hyderabad 500~032, India}

\author{Subhadip Mitra}
\email{subhadip.mitra@iiit.ac.in}
\affiliation{Center for Computational Natural Sciences and Bioinformatics, International Institute of Information Technology, Hyderabad 500~032, India}
\affiliation{Center for Quantum Science and Technology, \\International Institute of Information Technology, Hyderabad 500~032, India}

\begin{abstract}
Adding an entangled, non-interacting spectator to a two-body collision opens a window to study three-qubit correlations from a fundamental perspective. We examine seven representative tree-level quantum electrodynamics (QED) processes in this setup to investigate whether elementary scattering can generate genuine tripartite entanglement, distinguish between the inequivalent Greenberger-Horne-Zeilinger (GHZ) and W classes, and produce genuine tripartite nonlocality. Analysing the post-selected helicity states with the tangle, the generalised geometric measure, and the $\pi$-tangle reveals a kinematic inversion between a spectator entangled with one colliding particle and an initial GHZ state. Mass-induced helicity flips drive entanglement generation at low momentum at the cost of event rate, whereas helicity conservation protects an initial GHZ state at high momentum. Channel structure and selection rules also govern the entanglement class. Compton scattering and the Breit-Wheeler process generate exclusively W-class states with an identically vanishing tangle. For a GHZ input, photon-pair production converts the state into the W class at a scattering angle of $\pi/2$, while pure $s$-channel annihilation decouples the spectator completely. Finally, mapping a certified numerical lower bound on Svetlichny's inequality, benchmarked against an analytical upper bound, establishes genuine tripartite nonlocality in both setups.
\end{abstract}
\maketitle

\section{Introduction} \label{sec:intro}
\noindent
Multipartite entanglement is a key resource for distributed quantum information (QI) processing, with applications ranging from transferring quantum and classical information~\cite{yang2004efficient,srivastava2019oneshot,muhuri2026characterizing} to quantum secret sharing and key generation~\cite{singh2024controlled,abrol2025maximal,sazim2015retrieving}. Beyond two qubits, entanglement exhibits a richer structure. For example, a Greenberger-Horne-Zeilinger (GHZ) state cannot be converted into a W state, or vice versa, by local operations and classical communication (LOCC)~\cite{dur2000three}, even though both are three-qubit entangled states. A state can be entangled across every bipartition in more than one way. Genuine multipartite entanglement (GME) constrains the correlations and quantum information tasks achievable within the bipartite subsystems~\cite{prabhu2013exclusion,pandya2016complementarity}. It is also a resource in its own right~\cite{guhne2009entanglement}. Hence, a natural question is what physical interactions create such tripartite-entangled states.

We look for answers in particle scattering in quantum electrodynamics (QED), which determines how photons interact with electrically charged fermions, such as the electron.  When post-selected on a fixed scattering angle, the collision acts on the spin projections (helicities for fermions, polarisations for photons) of the colliding particles as a non-unitary two-qubit operation, determined by the rules of nature. If the particles could exchange a spinless mediator, the amplitude would factorise into separate operations on each particle, meaning such processes would not be entangling. However, the QED mediator is the spin-$1$ photon, which can create pairs of spin-entangled particles. Bipartite spin entanglement has been studied extensively in recent years, from Bell-inequality violations in QED scattering~\cite{manoukian2004speed,yongram2008spin,yongram2011spin} to the maximal entanglement of tree-level processes and its link to scattering cross sections~\cite{cervera-lierta2017maximal,fedida2023treelevel,blasone2025complete,seki2015variation,peschanski2016entanglement,fan2017variation,faleiro2020perturbative}. 
In this context, multipartite entanglement has been considered only recently, for instance in three-body decays~\cite{sakurai2024threebody}. Entanglement in multipartite systems has also been studied in consecutive scattering~\cite{quinta2024multipartite}, where multiple particles scatter sequentially off a single target; however, the study evaluates pairwise concurrence rather than GME.

A simple way to obtain a three-qubit state from a two-body scattering is to add a spectator, i.e., a third particle $C$ that is entangled with the colliding pair $A$ and $B$ beforehand but never takes part in the collision. In QI terms, $C$ plays the role of an ancilla, and the scattering acts on part of an entangled resource. Refs.~\cite{araujo2019measuring,fonseca2022entanglement,blasone2024entanglement,blasone2025entanglement,cao2026genuine} considered this configuration, mostly with $C$ entangled with one of the colliding particles, and followed the entropies of the subsystems and the bipartite correlations among $A$, $B$, and $C$ (Ref.~\cite{fonseca2022entanglement} also considered GHZ and W inputs). That, however, only shows a partial picture. Entropies and pairwise correlations miss the part of the three-qubit entanglement that no pair carries. This part can be maximal while every pairwise correlation vanishes. Hence, whether a collision entangles the three particles as a whole, and in which class, remains open. For a pure three-qubit state, the tangle measures the entanglement between one qubit and the other two that pairwise concurrences miss~\cite{coffman2000distributed}. It is maximal for the GHZ state but vanishes across the entire W class. The generalised geometric measure (GGM)~\cite{sende2010channel} and the $\pi$-tangle (the monogamy score of squared negativity)~\cite{ou2007monogamy,hamzehofi2026entanglement} detect GME in both classes, but no single measure distinguishes the two.

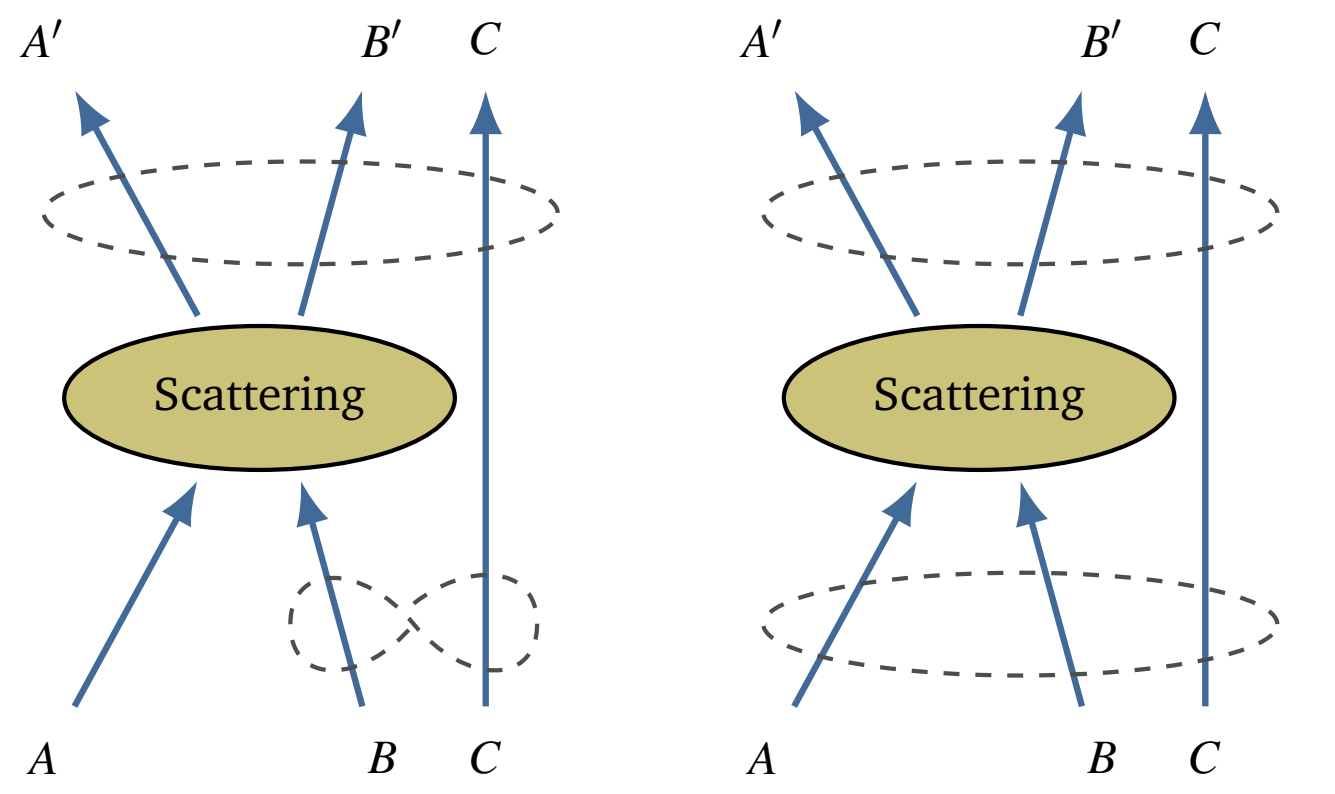
\begin{figure}[t]
\centering
\input{figures/qed_scattering}
\caption{The spectator configuration: Particles $A$ and $B$ scatter, while $C$ takes no part in the interaction. Dotted lines denote entanglement. The left figure shows $C$ entangled with $B$, as in the first input. In the GHZ input, $C$ is entangled with $A$ as well (right).}
\label{fig:system}
\end{figure}

We apply these three measures across various tree-level QED processes: M\o{}ller, Bhabha, Compton, and electron-muon scattering, muon-pair and photon-pair production, and the Breit-Wheeler process. Muons act as heavy electrons; their larger mass makes the helicity flips required to generate entanglement far more prominent. We consider two initial states. In the first, $C$ is entangled with $B$ alone while $A$ arrives uncorrelated, meaning any resulting GME must be generated by the collision itself. In the second, all three particles start in a GHZ state, testing whether the collision preserves existing entanglement. The outgoing GME depends strongly on the process and is largely dictated by angular-momentum selection rules. Compton scattering and the Breit-Wheeler process, for instance, produce only W-class entanglement and yield zero tangle across all kinematics. In contrast, annihilation through a single photon completely decouples the spectator from a GHZ input. Finally, we test the outgoing states for genuine tripartite nonlocality by mapping a certified numerical lower bound on the Svetlichny inequality~\cite{svetlichny1987distinguishing,werner2001bell,liang2007bounds} and benchmarking it against an analytical upper bound~\cite{li2017tight} at representative points. In a related study, Ref.~\cite{cao2026genuine} recently analysed genuine tripartite entanglement using the GGM, $\pi$-tangle, genuine multipartite concurrence, and concurrence fill, alongside the monogamy of squared entanglement of formation and squared quantum discord in Bhabha scattering with an uncorrelated-$A$ initial state. To our knowledge, however, there has not been any systematic study of GHZ/W classification, the survival of GME from an initial GHZ state, or genuine tripartite nonlocality across multiple QED processes in the presence of a spectator.

The plan for the paper is as follows. We set up the spectator configuration, the initial states, and the QED processes in Sec.~\ref{sec:preliminaries}; introduce the GME measures and the Svetlichny inequality in Sec.~\ref{sec:quantification}; study the creation and survival of GME and the selection rules behind them in Sec.~\ref{sec:results}; test the outgoing states for genuine tripartite nonlocality in Sec.~\ref{sec:svetlichny-results}; and conclude in Sec.~\ref{sec:conclusion}.

\section{Scattering with an entangled spectator} \label{sec:preliminaries}
\subsection{The setup} \label{sec:system}
\noindent
Let us consider two particles, $A$ and $B$, scattering in their centre-of-mass (COM) frame, alongside a third particle, $C$. The spectator $C$ is entangled either with $B$ or the pair beforehand but takes no part in the collision (Fig.~\ref{fig:system}), moving parallel to the incoming $B$ (say). We aim to determine the joint spin state of all three particles after the scattering event. Each particle acts as a single qubit representing its helicity (for fermions) or polarisation (for photons), labelled $\mathrm{P}$ (positive) or $\mathrm{N}$ (negative). We assume $C$ is sufficiently separated in phase space from $A$ and $B$ that it neither interacts with them nor requires antisymmetrisation if it shares their species.  

We use natural units and work entirely in the COM frame of $A$ and $B$. Here, the collision
\begin{equation}
A(p_1,s_1)+B(p_2,s_2)\longrightarrow A'(p_3,s_3)+B'(p_4,s_4)
\end{equation}
scatters the incoming particles, moving along the collision axis with three-momentum $\vert{}\momentum{p}\vert{}$, into an outgoing pair at a scattering angle $\theta\in[0,\pi]$. Here, $p_i$ are the four-momenta and $s_i\in\{{\mathrm{P}},{\mathrm{N}}\}$ the helicity labels. Particle masses set the scale at which helicity flips become important (Sec.~\ref{sec:qed-processes}), so we measure momentum in units of mass:
\begin{equation}
\xi_e=\frac{|\momentum{p}|}{m_e},\qquad \xi_\mu=\frac{|\momentum{p}|}{m_\mu},\qquad \lambda_m=\frac{m_e}{m_\mu},
\end{equation}
where $m_e$ and $m_\mu$ are the electron and muon masses. We use $\xi_\mu$ whenever a muon takes part in the collision, $\xi_e$ otherwise, and simply $\xi$ when the distinction does not matter. Also, we restrict our analysis to tree-level amplitudes. Higher-order effects, such as soft bremsstrahlung and loop corrections, are ignored, as they are parametrically suppressed in QED (unlike in strong interactions) and do not alter the fundamental helicity mappings.

The collision is governed by the S-matrix, $\mathcal{S}=\Id+i\mathcal{T}$, where $\mathcal{T}$ is the transition operator. Using covariantly normalised one-particle states, $\braket{\fermion{k}{a} \vert{} \fermion{p}{b}}=2E_{\momentum{k}}(2\pi)^3\delta^{(3)}(\momentum{k}-\momentum{p})\,\delta_{ab}$, where $E_{\momentum{k}}$ is the energy, we define the helicity amplitudes $\mathcal{M}_{s_1s_2;s_3s_4}$ through
\begin{align}
&\bra{\fermion{p_3}{s_3};\fermion{p_4}{s_4}} i \mathcal{T} \ket{\fermion{p_1}{s_1};\fermion{p_2}{s_2}} \nonumber\\
&\qquad= i(2\pi)^4\delta^{(4)}(p_1+p_2-p_3-p_4) \mathcal{M}_{s_1s_2;s_3s_4}.
\label{eq:T}
\end{align}
Because $C$ does not interact, the S-matrix acts on the three-particle state as $\mathcal{S}\otimes\Id_C$. By linearity, each helicity configuration of $AB$ scatters independently and carries its spectator helicity along. An initial helicity state $\ket{i}=\sum c_{s_1s_2s_C}\ket{s_1s_2}_{AB}\ket{s_C}_C$, with complex coefficients $c_{s_1s_2s_C}$ and spectator helicity $s_C$, therefore scatters into
\begin{align}
\ket{f}&=\sum_{s_1,s_2,s_C}c_{s_1s_2s_C}\,\ket{\phi_{s_1s_2}}_{AB}\ket{s_C}_C, \nonumber\\
\ket{\phi_{s_1s_2}}&=\sum_{s_3,s_4}\mathcal{M}_{s_1s_2;s_3s_4}\ket{s_3s_4}.
\label{eq:phi-branches}
\end{align}
Equation~\eqref{eq:phi-branches} isolates only the part of the state scattered into an angle $\theta$. Because a generic outgoing state contains all possible outcomes, restricting it to a fixed $\theta$ (which fixes the outgoing momenta via the delta function) and specific helicity qubits amounts to post-selecting on a scattering event~\cite{cervera-lierta2017maximal}. In matrix form, this transformation on the helicity subspace is $\ket{f}=(M\otimes\Id_C)\ket{i}$, where $M$ is the $4\times4$ amplitude matrix. Even though the full S-matrix is unitary on the complete spin-momentum Hilbert space, post-selecting on a fixed scattering angle makes the effective map $\ket{i}\to\ket{f}$ (and hence the amplitude matrix $M$) non-unitary. As a result, the unnormalised state $\ket{f}$ must be normalised as $\rho_{ABC}=\proj{f}/\braket{f|f}$, which also cancels the divergent overall factors from momentum conservation and avoids the need for an explicit regularisation scheme such as box regularisation~\cite{carney2016scattering,peschanski2019evaluation}.

Post-selecting events at a fixed angle acts on $A$ and $B$ alone. This makes it a local operation with respect to the $AB\vert{}C$ bipartition, meaning it cannot, on average, increase the entanglement across this split~\cite{bennett1996concentrating}. Consequently, large entanglement in the conditional state may come at the cost of a low event rate. For both inputs considered below, each spectator helicity tags a single incoming helicity configuration of $AB$. Because $\ket{\mathrm{P}}_C$ and $\ket{\mathrm{N}}_C$ are orthogonal, the unnormalised weight $\braket{f\vert{}f}$ reduces to an input-weighted sum of $\sum_{s_3s_4}\vert{}\mathcal{M}_{s_1s_2;s_3s_4}\vert{}^2$. At a fixed $(\xi,\theta)$, this weight is therefore proportional to the differential cross section of the prepared input, making ratios of $\braket{f\vert{}f}$ a direct measure of relative event rates.

\subsection{The initial states}\label{sec:initialstates}
\noindent
We define two specific initial states. 
\begin{enumerate}
    \item In the first, $A$ arrives with a definite helicity and $B$ and $C$ are entangled:
\begin{equation}
\ket{i}=\ket{{\mathrm{P}}}_A\otimes\left(\cos\eta\ket{{\mathrm{P}}{\mathrm{P}}}_{BC}+e^{i\beta}\sin\eta\ket{{\mathrm{N}}{\mathrm{N}}}_{BC}\right),
\label{eq:ispin}
\end{equation}
where $\eta\in[0,\pi/2]$ controls the $BC$ entanglement (ranging from a product state at $\eta=0$ to a Bell state at $\eta=\pi/4$), and $\beta$ is a relative phase. Because a local unitary on $C$ can absorb $\beta$, we set $\beta=0$ without loss of generality. In this case, any tripartite entanglement in the outgoing state must be generated by the collision.

\item 
In the second initial state, all three particles start in the maximally entangled GHZ state:
\begin{equation}
\ket{i}_{\text{GHZ}}=\frac{1}{\sqrt{2}}\left(\ket{{\mathrm{P}}{\mathrm{P}}{\mathrm{P}}}+\ket{{\mathrm{N}}{\mathrm{N}}{\mathrm{N}}}\right).
\label{eq:initial_ghz}
\end{equation}
Here, the relevant question is how much of this entanglement the collision preserves.
\end{enumerate}
For both initial states, applying Eq.~\eqref{eq:phi-branches} yields the following outgoing states:
\begin{align}
\ket{f}&=\cos\eta\,\ket{\phi_{{\mathrm{P}}{\mathrm{P}}}}_{AB}\ket{{\mathrm{P}}}_C+\sin\eta\,\ket{\phi_{{\mathrm{P}}{\mathrm{N}}}}_{AB}\ket{{\mathrm{N}}}_C, \label{eq:fspin}\\
\ket{f}&=\frac{1}{\sqrt{2}}\left(\ket{\phi_{{\mathrm{P}}{\mathrm{P}}}}_{AB}\ket{{\mathrm{P}}}_C+\ket{\phi_{{\mathrm{N}}{\mathrm{N}}}}_{AB}\ket{{\mathrm{N}}}_C\right), \label{eq:fghz}
\end{align}
which share a common structure,
\begin{equation}
\ket{f}=a\,\ket{\phi_x}_{AB}\ket{{\mathrm{P}}}_C+b\,\ket{\phi_y}_{AB}\ket{{\mathrm{N}}}_C.
\label{eq:generic-branches}
\end{equation}
The unnormalised states $\ket{\phi_x}$ and $\ket{\phi_y}$ act as distinct physical branches of the superposition. Each branch represents the complete outgoing state of $A$ and $B$ generated by scattering one specific incoming helicity configuration ($\mathrm{PP}$ or $\mathrm{PN}$ for the first initial state; $\mathrm{PP}$ or $\mathrm{NN}$ for the second). The helicity of $C$ acts as a tag, recording which configuration sourced which branch. Tripartite entanglement therefore depends entirely on how differently the collision treats these two incoming configurations.

The spectator $C$ decouples in two ways: if the collision treats the two configurations alike (making the branches parallel), or if one branch heavily suppresses the other ($\vert{}a\vert{}\lVert\phi_x\rVert\ll\vert{}b\vert{}\lVert\phi_y\rVert$ or vice versa, with $\lVert\cdot\rVert$ denoting the Euclidean norm). Genuine tripartite entanglement thus requires two non-parallel branches of comparable weight. For the first initial state, this immediately rules out GME at $\eta=0$ and $\eta=\pi/2$, where only a single branch survives.

\begin{table}[t]
\caption{\label{tab:processes}The seven tree-level QED processes and their contributing channels. The particles acting as $A$ and $B$ define the frame. Muon-pair production requires a centre-of-mass energy of at least $2m_\mu$, and the Breit-Wheeler process requires $\xi_e\geq1$. Processes with identical particles in the initial or final state are symmetric under $\theta\to\pi-\theta$.}
{\renewcommand{\arraystretch}{1.15}\small
\begin{tabular*}{\linewidth}{@{\extracolsep{\fill}}llll}
\hline\hline
Scattering process & $A$ & $B$ & Channels \\
\hline
M\o{}ller ($e^-e^-\to e^-e^-$) & $e^-$ & $e^-$ & $t,u$ \\
Photon-pair production ($e^-e^+\to\gamma\gamma$) & $e^-$ & $e^+$ & $t,u$ \\
Bhabha ($e^-e^+\to e^-e^+$) & $e^-$ & $e^+$ & $s,t$ \\
Muon-pair production ($e^-e^+\to\mu^-\mu^+$) & $e^-$ & $e^+$ & $s$ \\
Electron-muon ($\mu^-e^-\to\mu^-e^-$) & $\mu^-$ & $e^-$ & $t$ \\
Compton ($\gamma e^-\to\gamma e^-$) & $\gamma$ & $e^-$ & $s,u$ \\
Breit-Wheeler ($\gamma\gamma\to e^-e^+$) & $\gamma$ & $\gamma$ & $t,u$ \\
\hline\hline
\end{tabular*}}
\end{table}

\subsection{The QED processes} \label{sec:qed-processes}
\noindent
We evaluate the seven tree-level QED processes in Table~\ref{tab:processes}, covering photon exchange, fermion-pair annihilation, and external-photon scattering. Appendix~\ref{app:qed-helicity-amplitudes} details the Feynman amplitudes, spinor conventions, and resulting real-valued helicity amplitudes. Normalisation cancels the overall coupling factor $e^2$.

Two physical properties govern our analysis. First, QED vector coupling conserves helicity along a fermion line up to corrections of order $m/E$. Helicity flips therefore require slow-moving fermions, driving tripartite entanglement primarily at low momenta. Second, the channel structure dictates the branch weights. Pure $t$-channel processes preserve the incoming $A$ fermion line, whereas $s$- and $u$-channel diagrams break this continuity. In muon-pair production, for instance, every helicity configuration funnels through a single virtual photon.

\section{Quantifying entanglement and nonlocality}\label{sec:quantification}
\subsection{Genuine tripartite entanglement} \label{sec:genuine-tripartite-entanglement}
\noindent
A pure three-qubit state exhibits GME if it cannot be factorised across any of the bipartitions $A|BC$, $B|AC$, or $C|AB$. These states fall into two inequivalent classes, represented by $\ket{\text{GHZ}}=(\ket{000}+\ket{111})/\sqrt{2}$ and $\ket{\text{W}}=(\ket{001}+\ket{010}+\ket{100})/\sqrt{3}$~\cite{dur2000three}. A strict GME measure must vanish on all biseparable states and remain strictly positive on both classes. We evaluate the outgoing states using three complementary quantities:

\begin{itemize}
\item[--] \textbf{Tangle:} The tangle~\cite{coffman2000distributed} is defined as
\begin{equation}
\tau_{ABC}=\mathcal{C}^2_{A|BC}-\mathcal{C}^2_{AB}-\mathcal{C}^2_{AC}=4\left|\operatorname{Hdet}(\psi)\right|,
\label{eq:tangle}
\end{equation}
where $\mathcal{C}_{A|BC}=2\sqrt{\det\rho_A}$ is the concurrence between $A$ and the pair $BC$ (with $\rho_A$ the reduced state of $A$), and $\mathcal{C}_{AB}$ and $\mathcal{C}_{AC}$ are the two-qubit reduced concurrences. The term $\operatorname{Hdet}(\psi)$ is Cayley's hyperdeterminant of the state coefficients $\psi_{ijk}$ in the computational basis. The tangle is symmetric under qubit permutations and positive on the GHZ class, but it vanishes on the entire W class. Appendix~\ref{app:zero-tangle} provides a closed form for states of the branching structure in Eq.~\eqref{eq:generic-branches}.

\item[--] \textbf{$\pi$-tangle:} The $\pi$-tangle~\cite{ou2007monogamy} mirrors this construction but uses the negativity, $N_{X|Y}=\lVert\rho^{T_X}\rVert_1-1$, in place of concurrence. Here, $T_X$ denotes the partial transpose on $X$ and $\lVert\cdot\rVert_1$ the trace norm. (This normalisation equates the negativity and concurrence for pure two-qubit states.) Because the residual depends on the singled-out qubit, we average over all three choices:
\begin{align}
\pi_{ABC}&=\frac{1}{3}\left(\pi_A+\pi_B+\pi_C\right), \nonumber\\
\pi_A&=N^2_{A|BC}-N^2_{AB}-N^2_{AC},
\label{eq:pi-tangle}
\end{align}
with $\pi_B$ and $\pi_C$ defined similarly. Unlike the tangle, the $\pi$-tangle remains strictly positive on both classes.

\item[--] \textbf{Generalised geometric measure (GGM):} The GGM~\cite{sende2010channel} quantifies the distance to the nearest biseparable state. For pure three-qubit states, it reduces to
\begin{equation}
\mathcal{G}_{ABC}=1-\max\{\lambda_A,\lambda_B,\lambda_C\},
\label{eq:ggm}
\end{equation}
where $\lambda_X$ is the largest eigenvalue of the reduced state $\rho_X$. It is positive for all genuinely entangled pure states and bounded by $1/2$.
\end{itemize}

These measures distinguish the two classes. The GHZ state yields $\tau_{ABC}=\pi_{ABC}=1$ and $\mathcal{G}_{ABC}=1/2$, whereas the W state yields $\tau_{ABC}=0$, $\mathcal{G}_{ABC}=1/3$, and $\pi_{ABC}\approx0.55$. A pure state with zero tangle but non-zero GGM therefore belongs to the W class. 

\subsection{Svetlichny's inequality} \label{sec:svetlichny}
\noindent
Entanglement does not guarantee nonlocal correlations. To establish genuine tripartite nonlocality, correlations must defy any hybrid model where one party uses local hidden variables while the other two share arbitrary correlations~\cite{svetlichny1987distinguishing}. A violation of Svetlichny's inequality provides a sufficient test. If each party measures one of two spin observables -- for instance, $\hat A_i=\hat{\mathbf n}^A_i\cdot\boldsymbol\sigma$ for $A$, where $\hat{\mathbf n}^A_i$ is a unit vector and $\boldsymbol\sigma$ is the Pauli vector -- the Svetlichny operator is
\begin{align}
\hat{\mathcal S}_{\text{Sv}}={}&\hat A_1\big(\hat B_1\hat C_1+\hat B_1\hat C_2+\hat B_2\hat C_1-\hat B_2\hat C_2\big)\nonumber\\
&+\hat A_2\big(\hat B_1\hat C_1-\hat B_1\hat C_2-\hat B_2\hat C_1-\hat B_2\hat C_2\big).
\label{eq:svetlichny-operator}
\end{align}
Hybrid models obey $|\langle\hat{\mathcal S}_{\text{Sv}}\rangle|\leq4$, whereas quantum mechanics allows violations up to $4\sqrt2$.

Finding the maximum value $Q_{\text{Sv}}=\max|\mathrm{Tr}(\rho_{ABC}\hat{\mathcal S}_{\text{Sv}})|$ requires optimising over six unit vectors. Because a global search is impractical across full parameter-space scans, we compute a certified lower bound via multi-start see-saw optimisation~\cite{werner2001bell,liang2007bounds}. With two parties' axes fixed, the expectation is linear in the third party's measurement vectors, so each party's optimal axes follow in closed form. Sweeping sequentially through $A$, $B$, and $C$ cannot decrease the expectation. We run 40 sweeps from each of 48 random initialisations and take the largest final expectation $V_r$,
\begin{equation}
Q_{\text{Sv}}^{\text{LB}}=\max_{r=1,\ldots,48}V_r\leq Q_{\text{Sv}}.
\label{eq:svetlichny-seesaw-lb}
\end{equation}
Since each $V_r$ comes from explicit measurement settings, $Q_{\text{Sv}}^{\text{LB}}>4$ certifies genuine tripartite nonlocality. A value at or below $4$ remains inconclusive, as a local search cannot guarantee the global maximum.

To benchmark this lower bound at representative points, we pair it with the analytical upper bound of Ref.~\cite{li2017tight}. Unfolding the correlation tensor $t_{ijk}=\mathrm{Tr}[\rho_{ABC}\,\sigma_i\otimes\sigma_j\otimes\sigma_k]$ into a $3\times 9$ matrix $t^{(X)}$, with party $X$ labelling the rows, gives a valid bound for each party. The tightest of the three is
\begin{equation}
Q_{\text{Sv}}\leq Q_{\text{Sv}}^{\text{UB}}\equiv 4\min\left\{\lVert t^{(A)}\rVert_{\text{op}},\lVert t^{(B)}\rVert_{\text{op}},\lVert t^{(C)}\rVert_{\text{op}}\right\},
\label{eq:svetlichny-min-upper-bound}
\end{equation}
where $\lVert\cdot\rVert_{\text{op}}$ is the operator norm (the largest singular value). Equations~\eqref{eq:svetlichny-seesaw-lb} and~\eqref{eq:svetlichny-min-upper-bound} bracket $Q_{\text{Sv}}$ from both sides. Whenever $Q_{\text{Sv}}^{\text{LB}}$ meets $Q_{\text{Sv}}^{\text{UB}}$, the see-saw search has reached the global maximum.

\begin{figure*}[t]
\centering
\subfloat[Tangle, photon pair \label{fig:Tangle_photon-pair-production}]{%
    \includegraphics[width=0.24\linewidth]{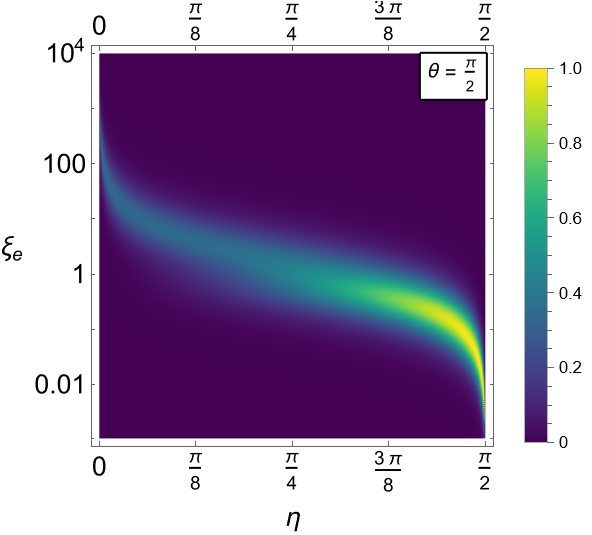}}\hfill
\subfloat[Tangle, M\o{}ller \label{fig:Tangle_moller}]{%
    \includegraphics[width=0.24\linewidth]{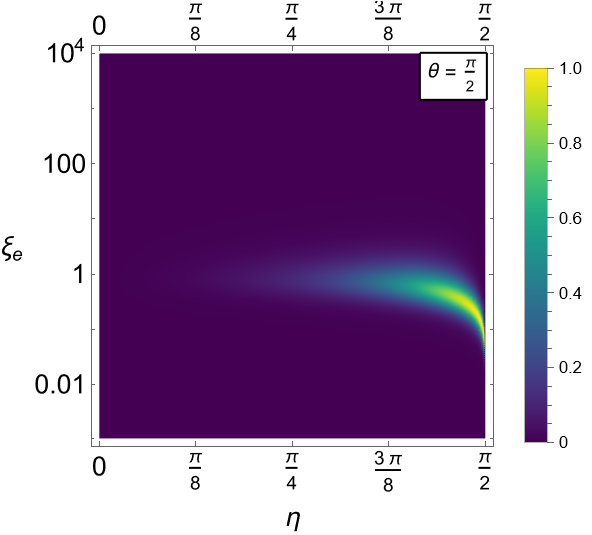}}\hfill
\subfloat[GGM, Breit-Wheeler \label{fig:GGM_breit-wheeler}]{%
    \includegraphics[width=0.24\linewidth]{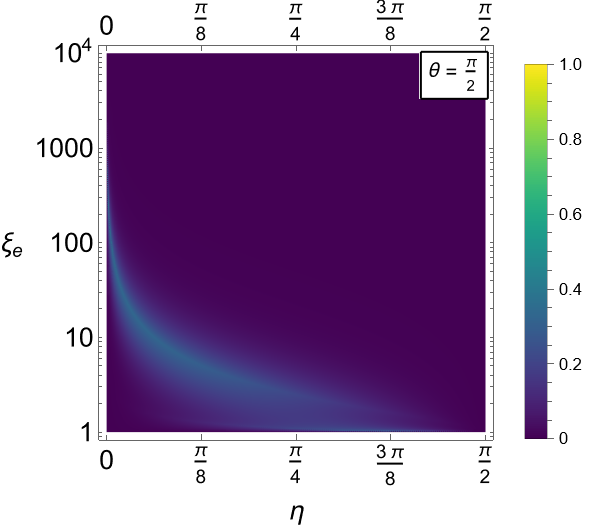}}\hfill
\subfloat[GGM, photon pair \label{fig:GGM_photon-pair-production}]{%
    \includegraphics[width=0.24\linewidth]{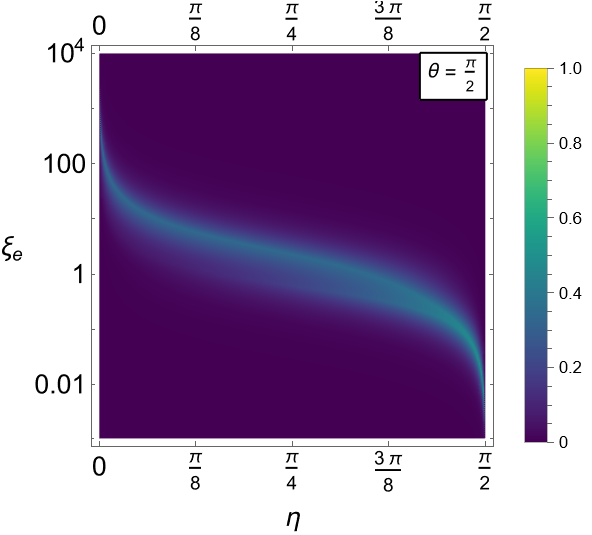}}\\
\subfloat[GGM, M\o{}ller \label{fig:GGM_moller}]{%
    \includegraphics[width=0.24\linewidth]{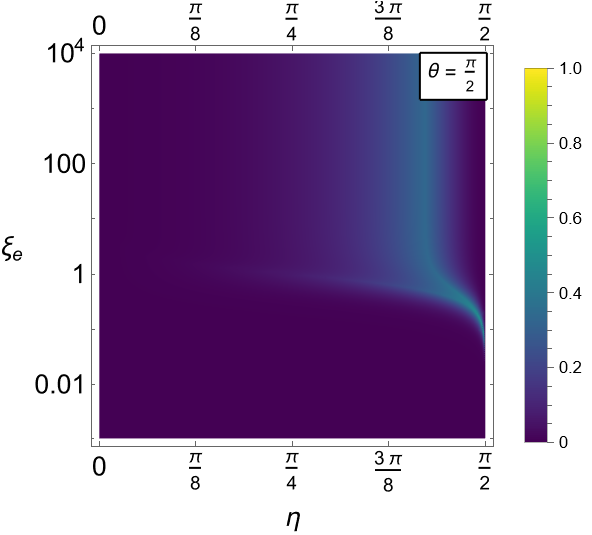}}\hfill
\subfloat[$\pi$-tangle, Breit-Wheeler \label{fig:pi-Tangle_breit-wheeler}]{%
    \includegraphics[width=0.24\linewidth]{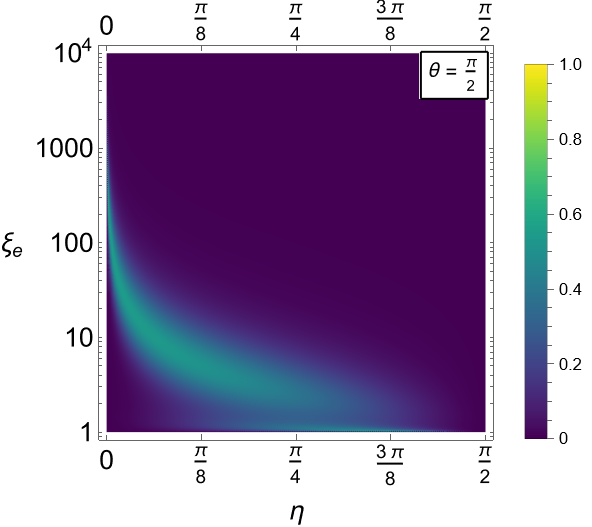}}\hfill
\subfloat[$\pi$-tangle, photon pair \label{fig:pi-Tangle_photon-pair-production}]{%
    \includegraphics[width=0.24\linewidth]{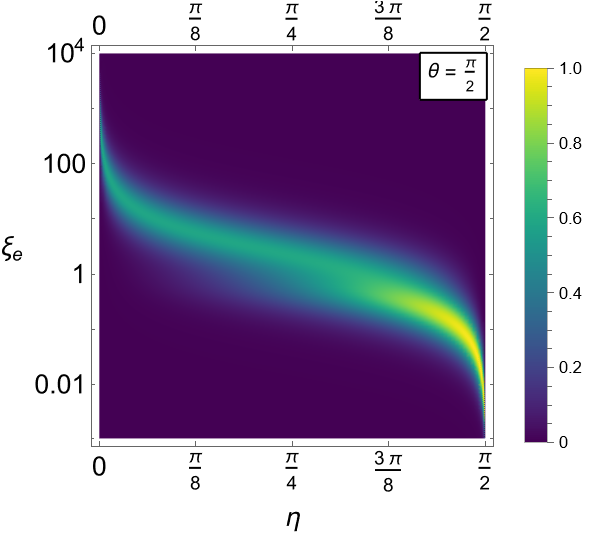}}\hfill
\subfloat[$\pi$-tangle, M\o{}ller \label{fig:pi-Tangle_moller}]{%
    \includegraphics[width=0.24\linewidth]{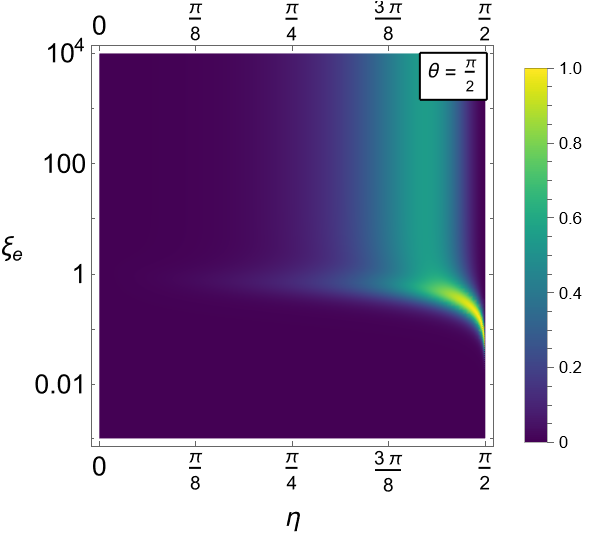}}
\caption{Tangle (a)--(b), GGM (c)--(e) and $\pi$-tangle (f)--(h) in the $(\xi_e,\eta)$ plane at $\theta=\pi/2$ for the processes symmetric under $\theta\to\pi-\theta$, with $A$ initially uncorrelated. The tangle of the Breit-Wheeler process vanishes and is not shown.}
\label{fig:symmetric-separable}
\end{figure*}
\begin{figure*}[t]
\centering
\subfloat[Electron-muon scattering \label{fig:Tangle_electron-muon}]{%
    \includegraphics[width=0.32\linewidth]{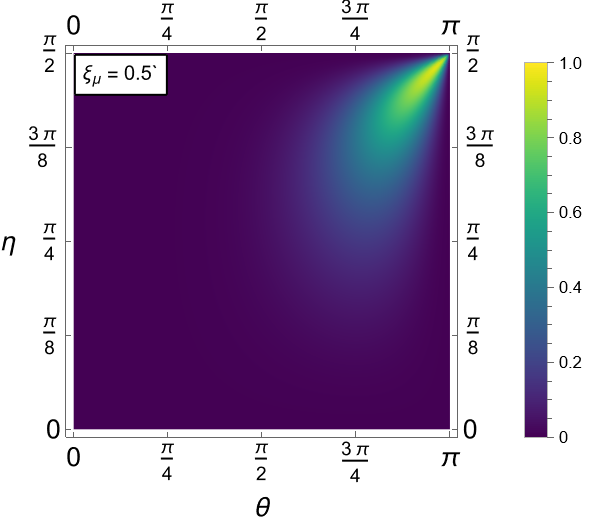}\hfill
    \includegraphics[width=0.32\linewidth]{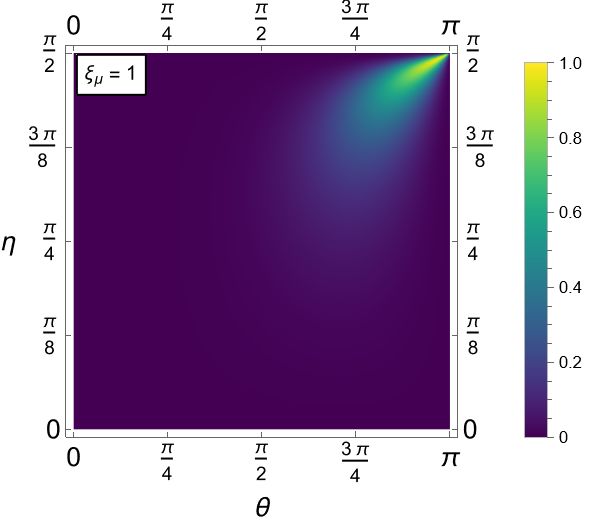}\hfill
    \includegraphics[width=0.32\linewidth]{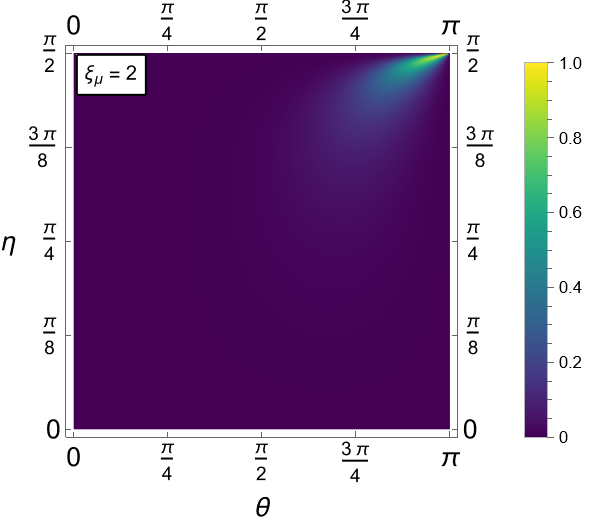}}\\
\subfloat[Bhabha scattering \label{fig:Tangle_bhabha}]{%
    \includegraphics[width=0.32\linewidth]{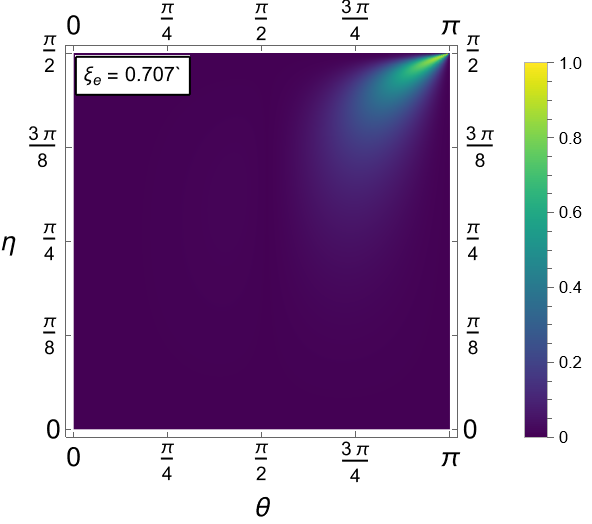}\hfill
    \includegraphics[width=0.32\linewidth]{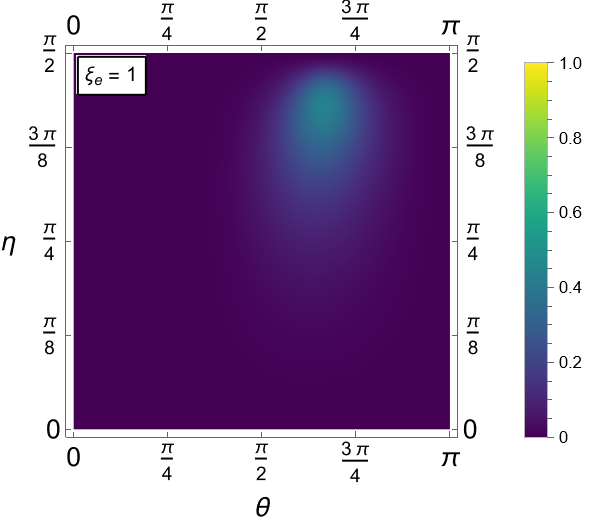}\hfill
    \includegraphics[width=0.32\linewidth]{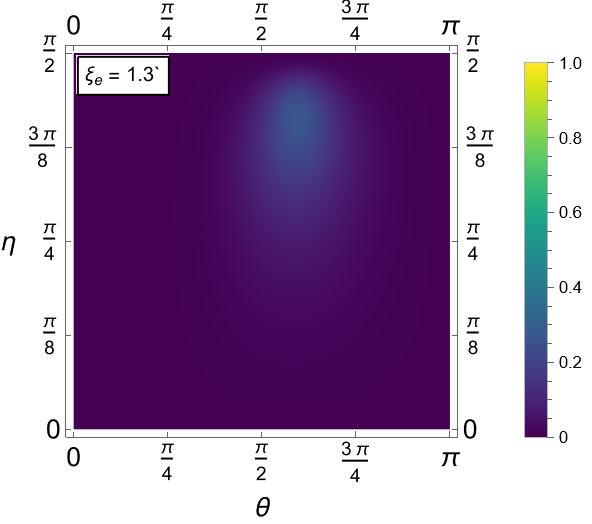}}
\caption{\label{fig:tangle-plots-for-bhabha-and-electron-muon-scattering} Tangle in the $(\eta,\theta)$ plane for (a) electron-muon and (b) Bhabha scattering, with $A$ initially uncorrelated, at the momenta given in each panel.}
\end{figure*}
\begin{figure*}[t]
\centering
\subfloat[Bhabha scattering \label{fig:GGM_bhabha}]{%
    \includegraphics[width=0.31\linewidth]{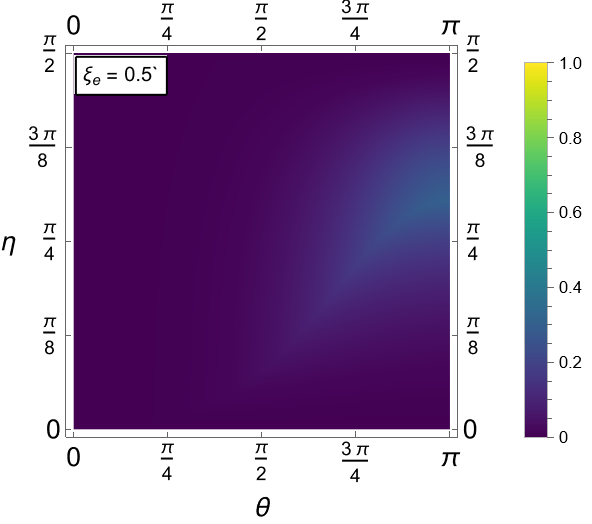}\hspace{0.01\linewidth}%
    \includegraphics[width=0.31\linewidth]{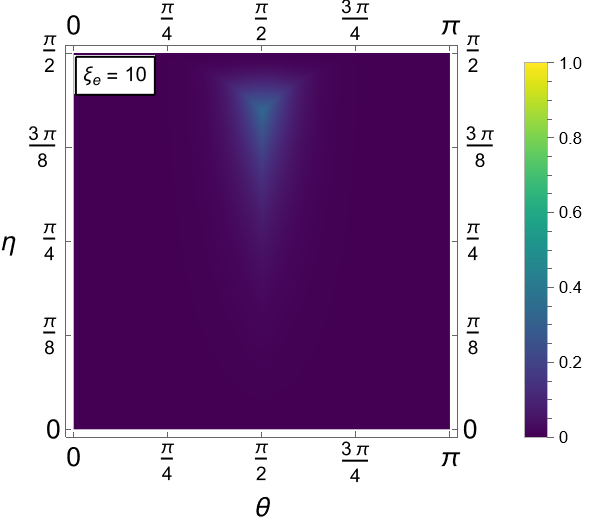}\hspace{0.01\linewidth}%
    \includegraphics[width=0.31\linewidth]{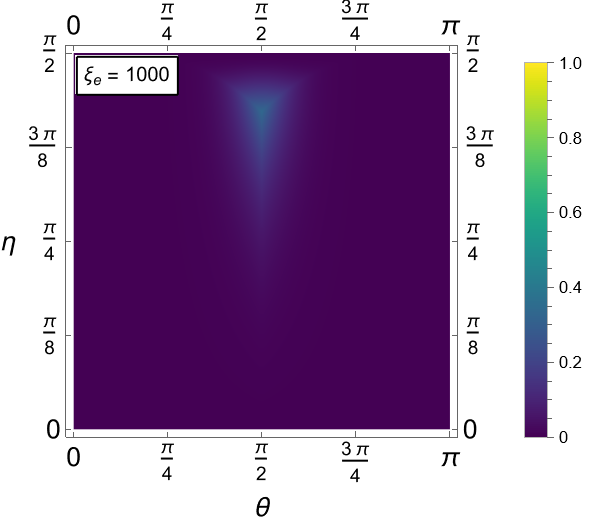}}\\
\subfloat[Compton scattering \label{fig:GGM_compton}]{%
    \includegraphics[width=0.31\linewidth]{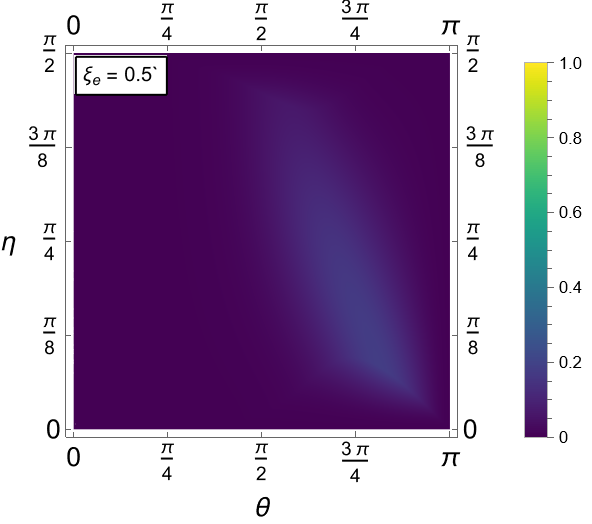}\hspace{0.01\linewidth}%
    \includegraphics[width=0.31\linewidth]{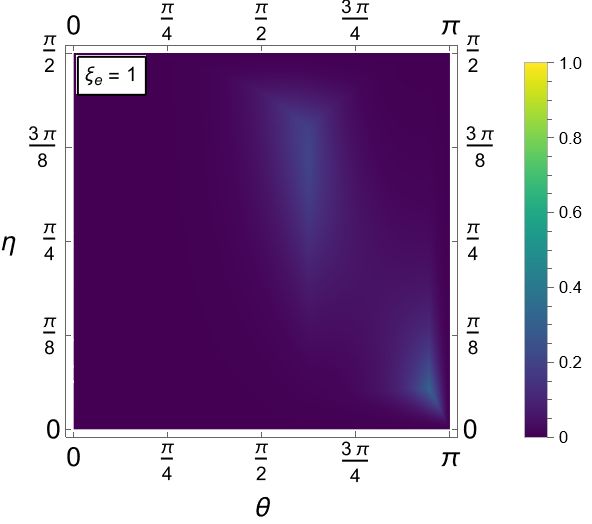}\hspace{0.01\linewidth}%
    \includegraphics[width=0.31\linewidth]{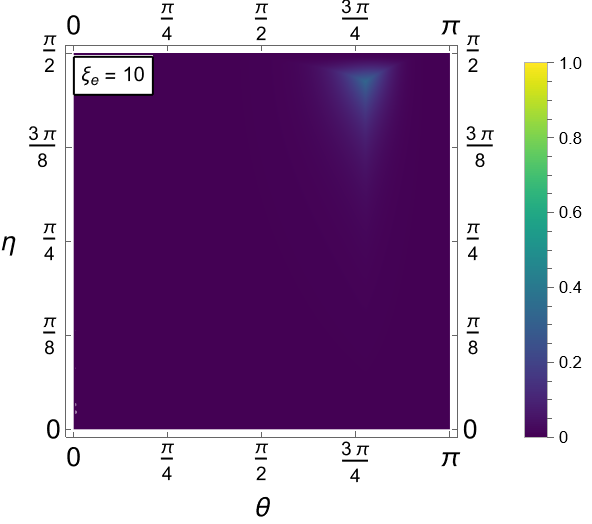}}\\
\subfloat[Electron-muon scattering \label{fig:GGM_electron-muon}]{%
    \includegraphics[width=0.31\linewidth]{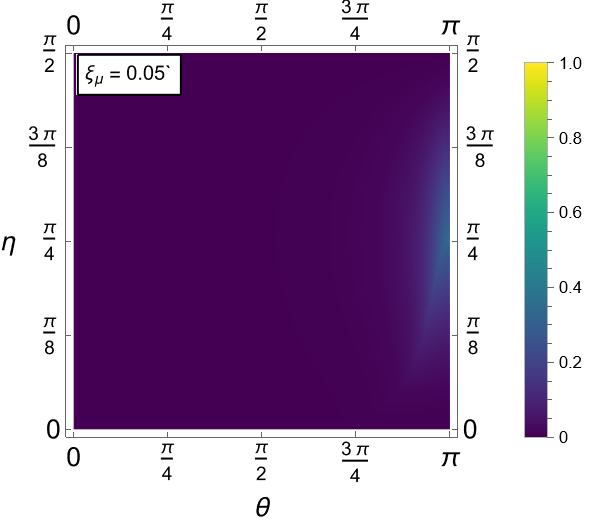}\hspace{0.01\linewidth}%
    \includegraphics[width=0.31\linewidth]{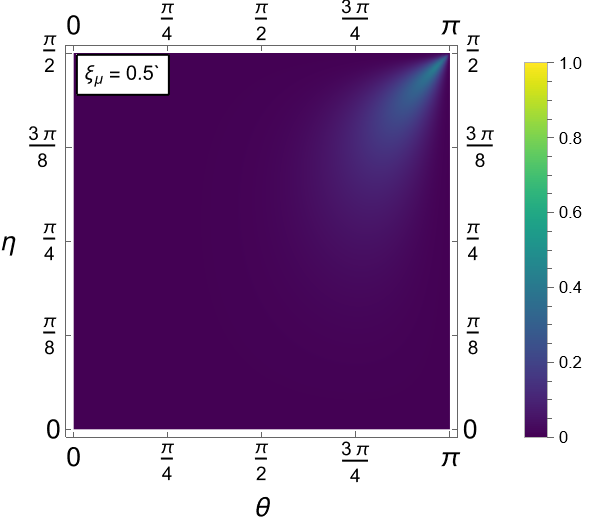}\hspace{0.01\linewidth}%
    \includegraphics[width=0.31\linewidth]{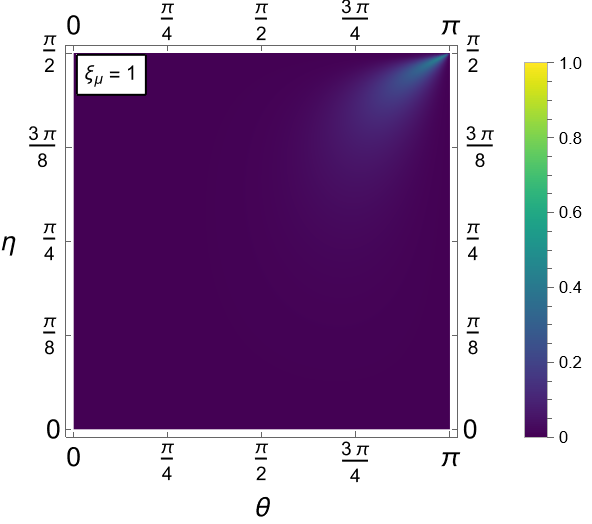}}
\caption{\label{fig:ggm-plots-for-bhabha-and-compton-and-electron-muon-scattering} GGM in the $(\eta,\theta)$ plane for (a) Bhabha, (b) Compton and (c) electron-muon scattering, with $A$ initially uncorrelated, at the momenta given in each panel.}
\end{figure*}
\begin{figure*}[t]
\centering
\subfloat[Bhabha scattering \label{fig:pi-Tangle_bhabha}]{%
    \includegraphics[width=0.31\linewidth]{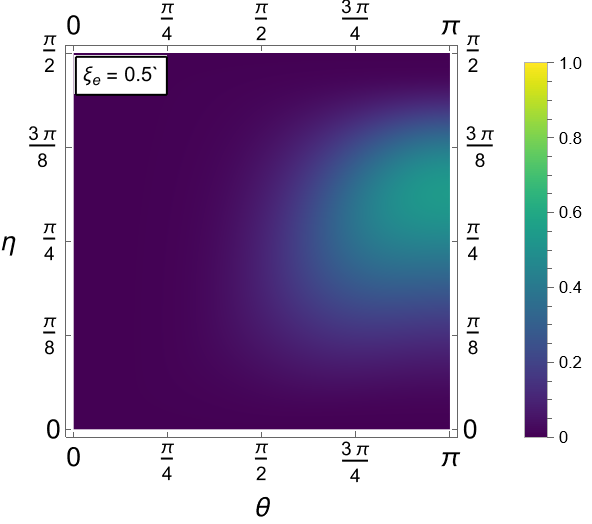}\hspace{0.01\linewidth}%
    \includegraphics[width=0.31\linewidth]{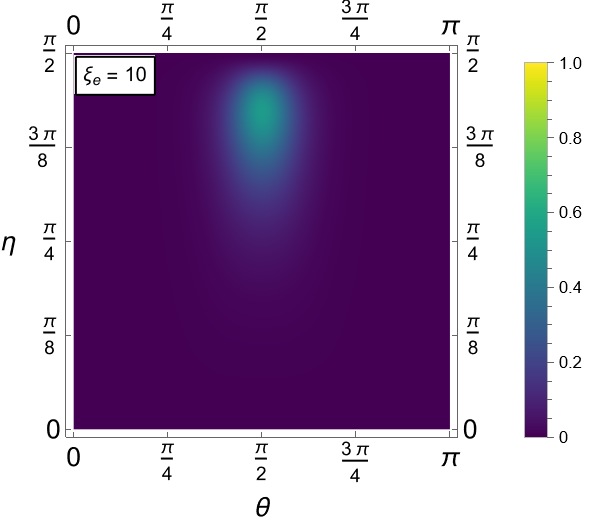}\hspace{0.01\linewidth}%
    \includegraphics[width=0.31\linewidth]{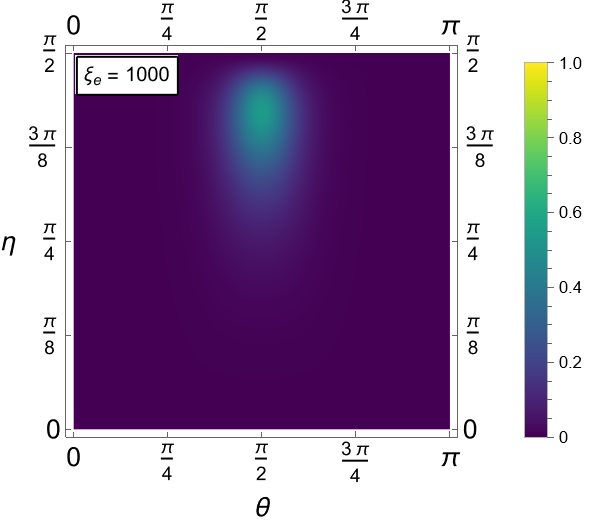}}\\
\subfloat[Compton scattering \label{fig:pi-Tangle_compton}]{%
    \includegraphics[width=0.31\linewidth]{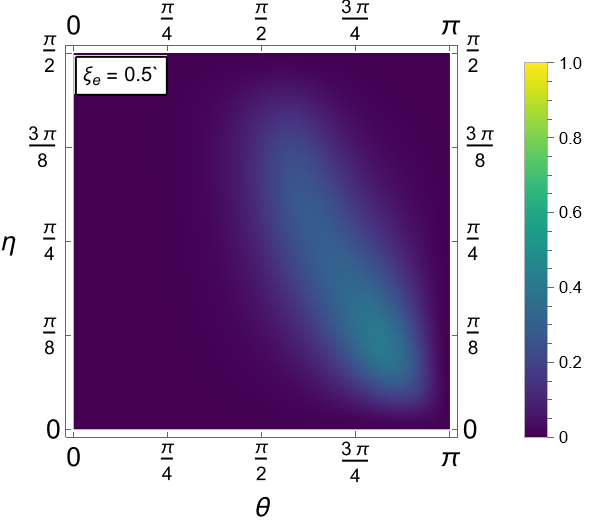}\hspace{0.01\linewidth}%
    \includegraphics[width=0.31\linewidth]{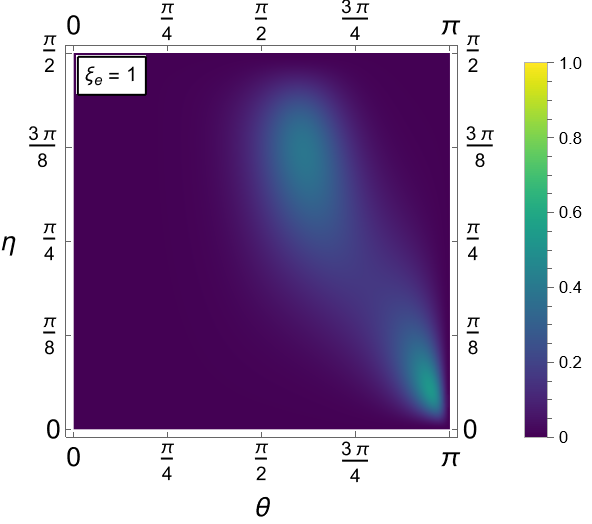}\hspace{0.01\linewidth}%
    \includegraphics[width=0.31\linewidth]{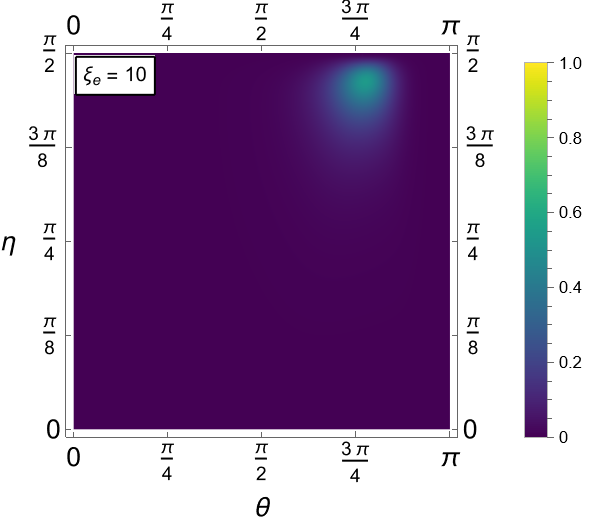}}\\
\subfloat[Electron-muon scattering \label{fig:pi-Tangle_electron-muon}]{%
    \includegraphics[width=0.31\linewidth]{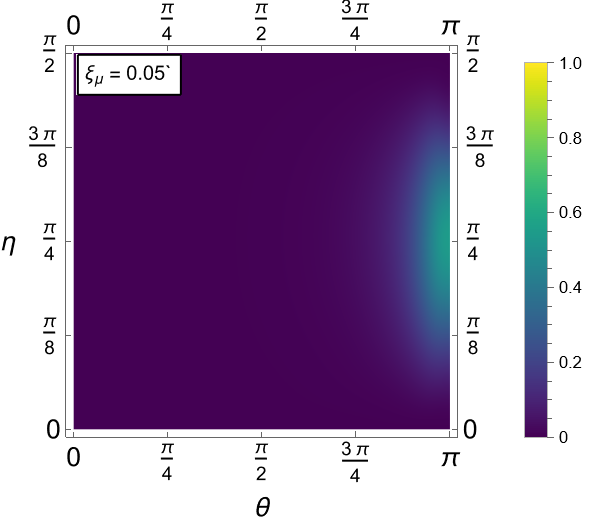}\hspace{0.01\linewidth}%
    \includegraphics[width=0.31\linewidth]{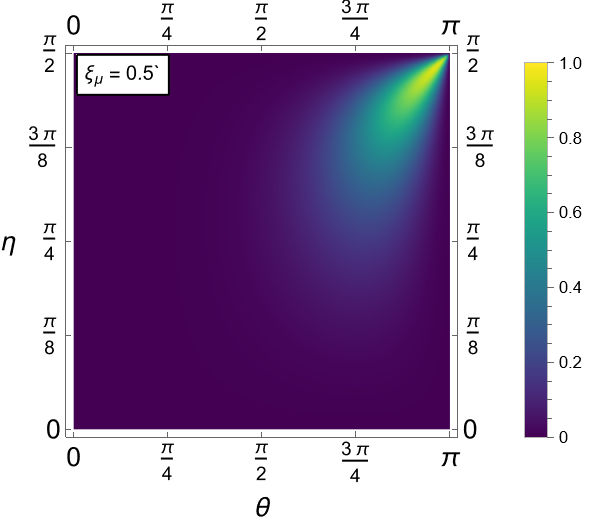}\hspace{0.01\linewidth}%
    \includegraphics[width=0.31\linewidth]{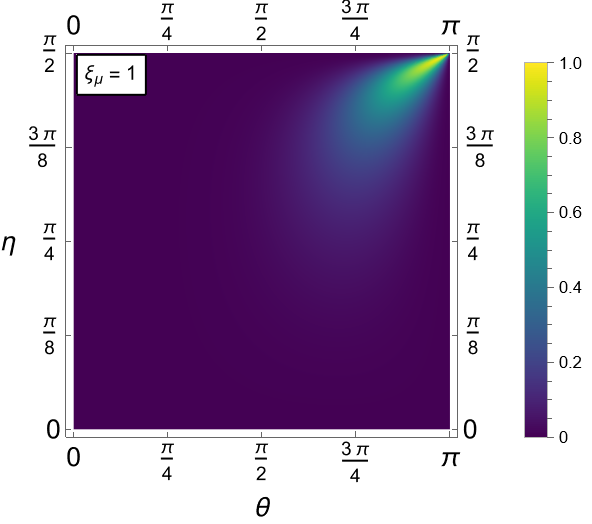}}
\caption{\label{fig:pi-tangle-plots-for-bhabha-and-compton-and-electron-muon-scattering} $\pi$-tangle in the $(\eta,\theta)$ plane for (a) Bhabha, (b) Compton and (c) electron-muon scattering, with $A$ initially uncorrelated, at the momenta given in each panel.}
\end{figure*}
\begin{figure*}[t]
\centering
\subfloat[Tangle, muon pair \label{fig:muon-pair-separable-tangle}]{%
    \includegraphics[width=0.32\linewidth]{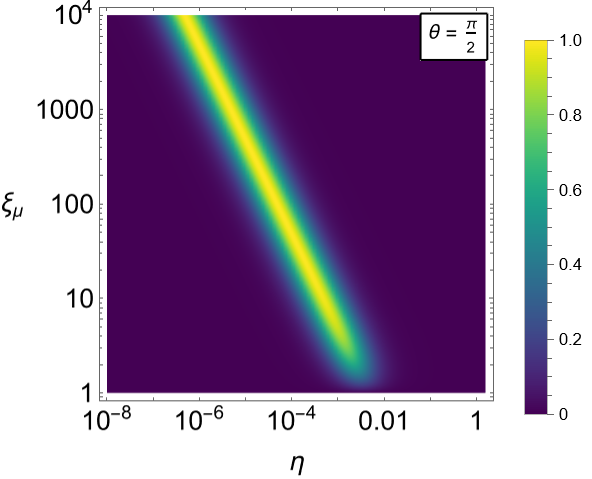}}\hfill
\subfloat[GGM, muon pair \label{fig:muon-pair-separable-ggm}]{%
    \includegraphics[width=0.32\linewidth]{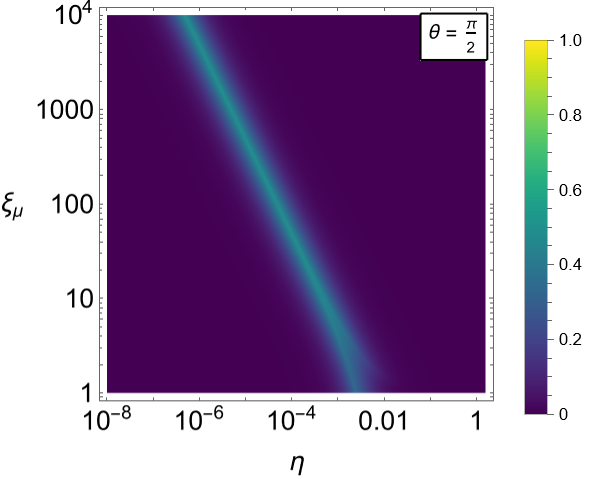}}\hfill
\subfloat[$\pi$-tangle, muon pair \label{fig:muon-pair-separable-pi-tangle}]{%
    \includegraphics[width=0.32\linewidth]{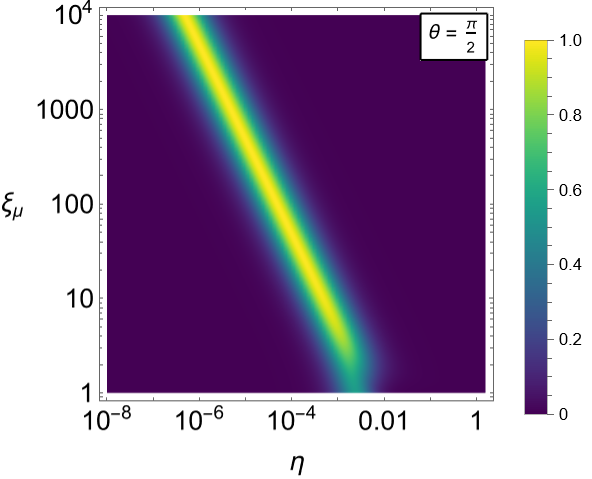}}
\caption{\label{fig:muon-pair-separable-measures} The three measures for muon-pair production with $A$ initially uncorrelated, in the $(\eta,\xi_\mu)$ plane at $\theta=\pi/2$. The logarithmic scale in $\eta$ resolves the narrow ridge at small $\eta$.}
\end{figure*}

\section{Post-scattering genuine tripartite entanglement} \label{sec:results}
\noindent
In the COM frame, an incoming pair with helicities $\lambda_1$ and $\lambda_2$ carries angular momentum $J_z=\lambda_1-\lambda_2$ along the collision axis~\cite{jacob1959general}. Each branch of Eq.~\eqref{eq:generic-branches} grows from one incoming helicity configuration and therefore carries a definite $J_z$. Angular-momentum conservation and the chirality of QED dictate whether the collision suppresses one branch or aligns the two (the two decoupling mechanisms identified in Sec.~\ref{sec:initialstates}). Because helicity flips cost an amplitude factor of order $m/E$, they provide the primary mechanism to differentiate the branches. Tripartite entanglement thus depends strongly on the particle masses and the collision momentum.

\subsection{Generation from an uncorrelated $A$} \label{sec:tripartite-correlation-results}
\noindent
Figures~\ref{fig:symmetric-separable} to~\ref{fig:muon-pair-separable-measures} map the three entanglement measures across parameter space for the various processes, omitting those that are negligible everywhere. We plot the symmetric processes at $\theta=\pi/2$ (because the GGM and $\pi$-tangle can peak away from this central angle at certain momenta, these slices represent a conservative estimate of the total GME). Three broad features emerge: first, GME peaks at low momentum, where helicity flips are least suppressed because masses are not yet negligible compared to the momentum scale; second, the tangle is highly restrictive (vanishing entirely in Compton scattering and the Breit-Wheeler process); and third, the GGM and $\pi$-tangle survive into the relativistic regime. We can trace these features to specific kinematic mechanisms.

In M\o{}ller scattering and photon-pair production, the tangle maximises only along a narrow curved ridge approaching $\eta=\pi/2$ at low momentum [Figs.~\ref{fig:Tangle_photon-pair-production} and~\ref{fig:Tangle_moller}], but remains small for a balanced input ($\eta=\pi/4$) despite maximal initial $BC$ entanglement. Branch imbalance explains this. At low momentum, the collision heavily suppresses the $\mathrm{PN}$ branch. In M\o{}ller scattering, the $t$- and $u$-channel terms cancel each other at $\theta=\pi/2$, leaving $\lVert\phi_{\mathrm{PP}}\rVert/\lVert\phi_{\mathrm{PN}}\rVert\approx1/\xi_e^2$. In photon-pair production, the Landau-Yang theorem suppresses it, yielding a ratio of $\approx1/\xi_e$. Generating a large tangle therefore requires an unbalanced input ($\tan\eta \gg 1$) to compensate, pushing the ridge towards $\eta=\pi/2$ as $\xi_e$ decreases. Because the input places almost all its weight on the suppressed branch, these high-entanglement events suffer a severely reduced event rate.

In electron-muon scattering [Fig.~\ref{fig:Tangle_electron-muon}], the tangle is large when the muon is mildly relativistic ($\xi_\mu \sim 1$) but retreats to a shrinking corner ($\theta\to\pi$, $\eta\to\pi/2$) as momentum grows. In a pure $t$-channel process, the outgoing $A$ continues the incoming $A$ line. For the first input, $A$ arrives with helicity $\mathrm{P}$. If it retained this helicity across all branches, it would factor out of Eq.~\eqref{eq:fspin}. Preventing this requires a mass-induced flip proportional to $m_A/E_A$. Generating entanglement therefore requires the uncorrelated particle to be slow enough to flip. When $A$ is the muon, this flip is appreciable for $\xi_\mu \sim 1$, whereas the electron is already ultrarelativistic. If $A$ is the electron, both the flip probability and the resulting tangle drop by several orders of magnitude. When $u$- or $s$-channel diagrams contribute, as in Bhabha scattering, the outgoing particle at angle $\theta$ need not continue the incoming $A$ line, confining the tangle to a narrow band around $\xi_e\approx1$ [Fig.~\ref{fig:Tangle_bhabha}].

In muon-pair production, a narrow straight ridge appears at small $\eta$ on a logarithmic scale [Fig.~\ref{fig:muon-pair-separable-measures}], while the measures vanish elsewhere. This occurs because the amplitude factorises through a single virtual photon:
\begin{equation}
\mathcal{M}_{s_1s_2;s_3s_4}=\frac{e^2}{s}\,J_{\text{in}}^{\mu}(s_1s_2)\,J^{\text{out}}_{\mu}(s_3s_4),
\label{eq:s-channel-factorisation}
\end{equation}
with currents $J_{\text{in}}^\mu=\bar v_2\gamma^\mu u_1$ and $J^{\text{out}}_\mu=\bar u_3\gamma_\mu v_4$, and $s=(p_1+p_2)^2$. In the COM frame, the virtual photon has zero spatial momentum; current conservation ($q_\mu J_{\text{in}}^\mu = 0$) therefore forces the temporal component to vanish, leaving the incoming current as a pure 3-vector ($J_{\text{in}}^0=0$). 

The direction of this 3-vector completely determines the shape of the outgoing branch. Its direction, in turn, is dictated by the incoming helicities. If the incoming pair has the same helicity ($J_z=0$), their spins are antiparallel, forcing the current vector to point strictly along the collision axis ($\bm{J}_{\text{in}}\propto\hat{\bm z}$). If they have opposite helicities ($J_z=\pm1$), their spins align, forcing the current into the transverse plane along the polarisation vector ($\bm{J}_{\text{in}}\propto\hat{\bm\epsilon}_\pm$). By substituting Eq.~\eqref{eq:s-channel-factorisation} into the branch definition of Eq.~\eqref{eq:phi-branches}, every branch becomes proportional to a fixed state dictated by this specific polarisation:
\begin{equation}
\ket{\phi_{s_1s_2}}\propto\ket{\chi_{J_z}}=\sum_{s_3,s_4}\hat{\bm e}^*_{J_z}\cdot\bm{J}_{\text{out}}(s_3s_4)\ket{s_3s_4},
\end{equation}
with $\hat{\bm e}_0=\hat{\bm z}$ and $\hat{\bm e}_{\pm1}=\hat{\bm\epsilon}_\pm$.

For the first input, the $\mathrm{PP}$ branch stems from the same-helicity configuration, while the $\mathrm{PN}$ branch stems from the opposite-helicity configuration. Because they couple to orthogonal polarisations of the virtual photon, the two branches cannot align. They are, however, severely unbalanced. Annihilating a same-helicity pair requires a mass-induced helicity flip, so $\lVert\phi_{\mathrm{PP}}\rVert/\lVert\phi_{\mathrm{PN}}\rVert\approx\varepsilon$, where $\varepsilon\sim m_e/E_e\leq\lambda_m$. The branch weights in Eq.~\eqref{eq:generic-branches} therefore scale as $\varepsilon\cot\eta$. For generic $\eta$, the $\mathrm{PN}$ branch dominates and $C$ decouples. The branches balance only when $\tan\eta\approx\varepsilon$, creating a narrow tripartite-entanglement window at $\eta\sim\tan^{-1}(m_e/E_e)$. Along this ridge, the GGM approaches its theoretical maximum of $1/2$, though the event rate drops by a factor of roughly $\eta^2$.

Ultimately, for an uncorrelated $A$ input, the collision generates GME wherever it treats the two helicities of $B$ differently while keeping both branches comparable in weight. In QED, this requires helicity flips at low momentum, demanding an unbalanced input to maximise the entanglement. In Compton scattering and the Breit-Wheeler process, the branches are neither negligible nor parallel. The outgoing state thus carries GME, reflected by a positive GGM and $\pi$-tangle [Figs.~\ref{fig:GGM_breit-wheeler}, \ref{fig:GGM_compton}, \ref{fig:pi-Tangle_breit-wheeler}, and~\ref{fig:pi-Tangle_compton}]. Despite this, the tangle vanishes identically across all kinematics, leading to the following exact classification:

\medskip
\noindent\textbf{Statement 1.} \textit{For an uncorrelated $A$ input, the tangle of the outgoing state vanishes identically in Compton scattering and the Breit-Wheeler process across all kinematics. Wherever the outgoing state carries GME, it strictly belongs to the W class.}
\smallskip

\noindent\textit{Proof.} We show in Appendix~\ref{app:zero-tangle} that the tangle of a two-branch state is proportional to the discriminant of a quadratic polynomial built from the branches. For both processes, this discriminant cancels. Because every pure state with GME outside the W class has a positive tangle~\cite{dur2000three}, the state must belong to the W class. \hfill$\square$

\begin{figure*}[t]
\centering
\subfloat[Bhabha scattering \label{fig:Tangle_bhabha-initial-ghz}]{%
    \includegraphics[width=0.31\linewidth]{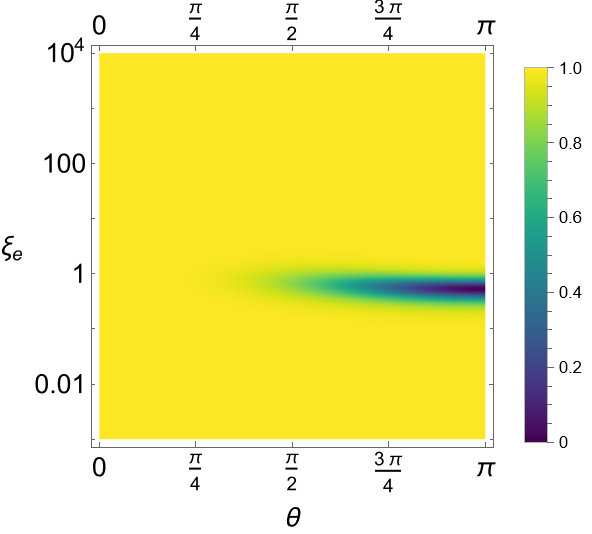}}\hfill
\subfloat[Breit-Wheeler process \label{fig:Tangle_breit-wheeler-initial-ghz}]{%
    \includegraphics[width=0.31\linewidth]{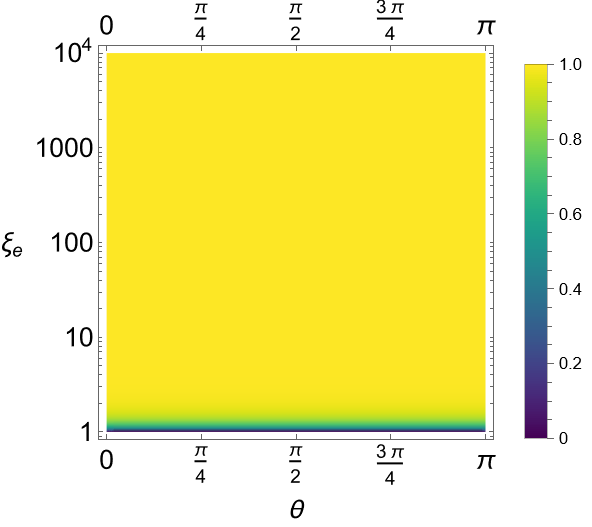}}\hfill
\subfloat[Electron-muon scattering \label{fig:Tangle_electron-muon-initial-ghz}]{%
    \includegraphics[width=0.31\linewidth]{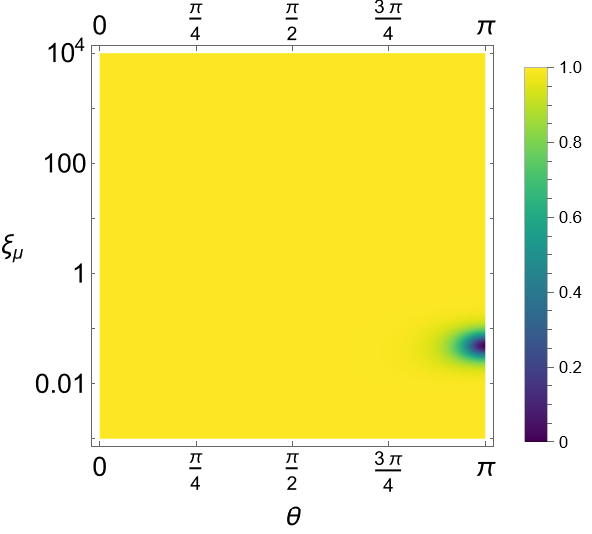}}\\
\subfloat[M\o{}ller scattering \label{fig:Tangle_moller-initial-ghz}]{%
    \includegraphics[width=0.31\linewidth]{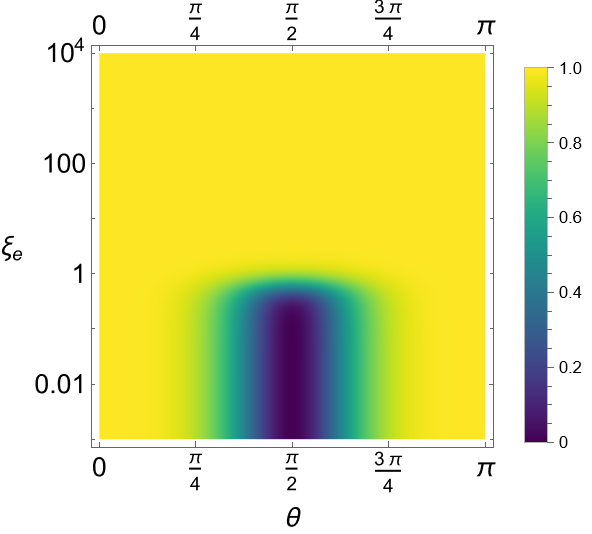}}\hfill
\subfloat[Compton scattering \label{fig:Tangle_compton-initial-ghz}]{%
    \includegraphics[width=0.31\linewidth]{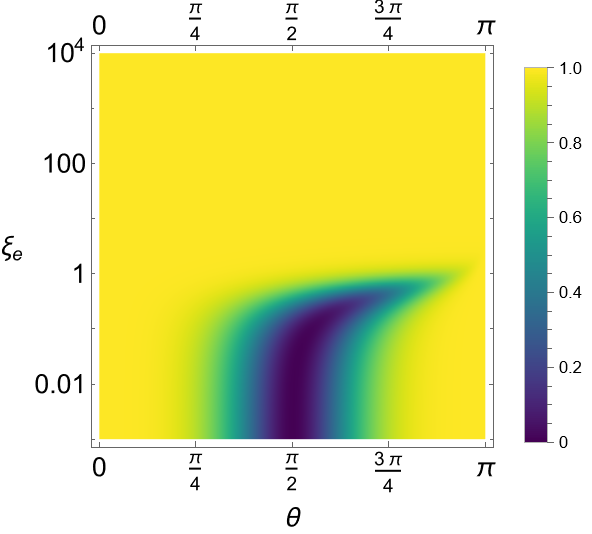}}\hfill
\subfloat[Photon-pair production \label{fig:Tangle_photon-pair-production-initial-ghz}]{%
    \includegraphics[width=0.31\linewidth]{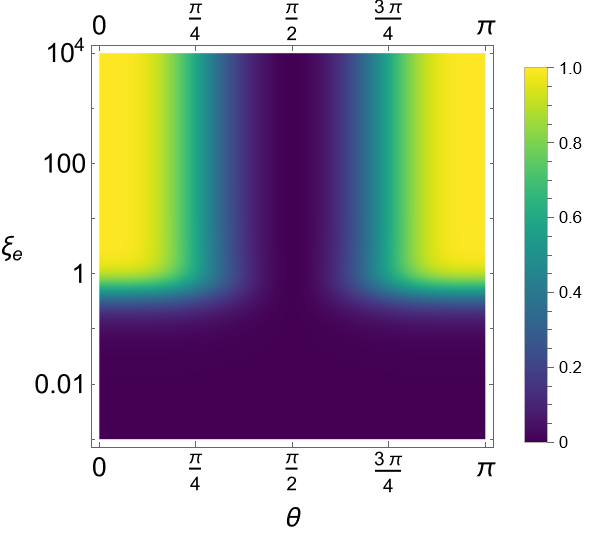}}
\caption{\label{fig:ghz_tangle} Tangle in the $(\theta,\xi)$ plane for an initial GHZ state. Top row: processes that preserve near-maximal entanglement over almost all of phase space. Bottom row: processes with sizeable regions of suppression. Muon-pair production gives zero and is not shown.}
\end{figure*}
\begin{figure}[t]
\centering
\subfloat[GGM \label{fig:GGM_photon-pair-production-initial-ghz}]{%
    \includegraphics[width=0.6\columnwidth]{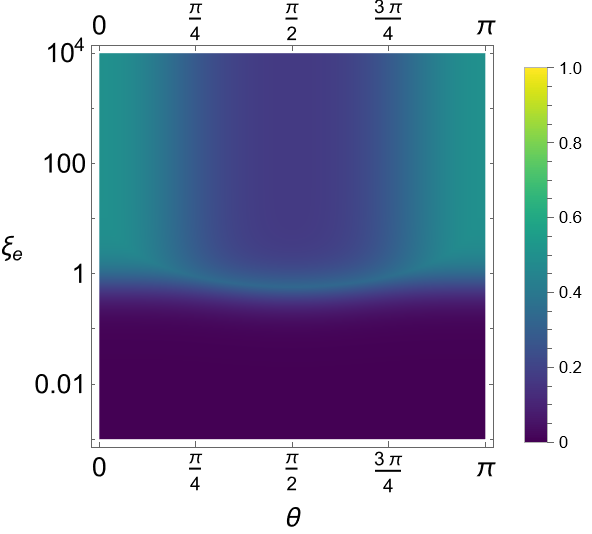}}\\
\subfloat[$\pi$-tangle \label{fig:piTangle_photon-pair-production-initial-ghz}]{%
    \includegraphics[width=0.6\columnwidth]{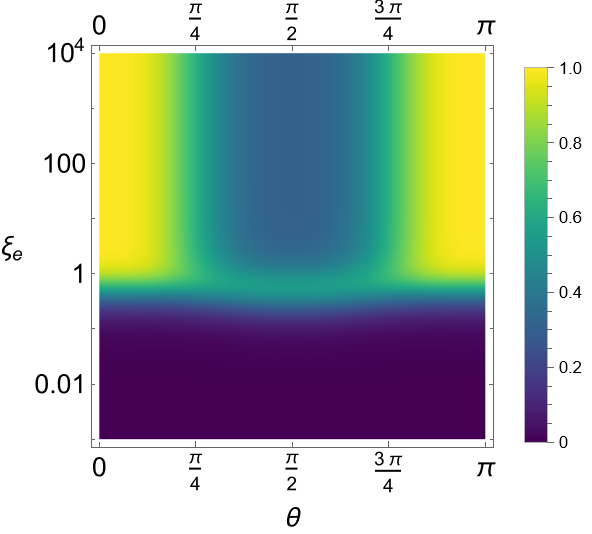}}
\caption{\label{fig:ghz-photon-pair} GGM and $\pi$-tangle in the $(\theta,\xi_e)$ plane for photon-pair production from an initial GHZ state. Both stay positive at $\theta=\pi/2$, where the tangle vanishes.}
\end{figure}

\subsection{Survival of an initial GHZ state} \label{sec:results-initial-ghz}
\noindent
For the GHZ initial state [Eq.~\eqref{eq:initial_ghz}], the branches $\ket{\phi_{\mathrm{PP}}}$ and $\ket{\phi_{\mathrm{NN}}}$ both carry $J_z=0$. Preserving its maximal entanglement requires every single-qubit reduced state to remain maximally mixed. For $C$, this demands that two branches of equal norm remain strictly orthogonal. The monogamy relation $\tau_{ABC}\leq\mathcal{C}^2_{C|AB}=4\det\rho_C$ quantifies this constraint:
\begin{equation}
\tau_{ABC}\leq 1-\frac{|\braket{\phi_{\mathrm{PP}}|\phi_{\mathrm{NN}}}|^2}{\lVert\phi_{\mathrm{PP}}\rVert^2\,\lVert\phi_{\mathrm{NN}}\rVert^2}.
\label{eq:ghz-bound}
\end{equation}
The second term is the squared overlap of the normalised branches. Orthogonal branches preserve the full tangle, parallel branches destroy it entirely, and the collision degrades the GHZ entanglement proportionally to the branch overlap it induces.

Figure~\ref{fig:ghz_tangle} maps the tangle for the six processes where it does not vanish identically. The $\pi$-tangle visually coincides with the tangle for every process except photon-pair production, and the GGM follows the identical structural pattern (see Appendix~\ref{app:zero-tangle}). For photon-pair production, Fig.~\ref{fig:ghz-photon-pair} maps the GGM and $\pi$-tangle explicitly. The Breit-Wheeler process, Bhabha scattering, and electron-muon scattering preserve near-maximal entanglement over almost the entire phase space [Figs.~\ref{fig:Tangle_bhabha-initial-ghz}--\ref{fig:Tangle_electron-muon-initial-ghz}]. In the Breit-Wheeler process, the photon pair with $J_z=0$ produces only equal-helicity fermion pairs at tree level (see Appendix~\ref{app:qed-helicity-amplitudes}):
\begin{align}
\ket{\phi_{\mathrm{PP}}}&=a_+\ket{\mathrm{PP}}+a_-\ket{\mathrm{NN}}, \nonumber\\
\ket{\phi_{\mathrm{NN}}}&=-a_-\ket{\mathrm{PP}}-a_+\ket{\mathrm{NN}},
\end{align}
with $a_\pm\propto\xi_e\pm\sqrt{\xi_e^2-1}$ sharing a $\theta$-dependent common factor. This structure leads directly to an exact kinematic constraint on the surviving tangle:

\medskip
\noindent\textbf{Statement 2.} \textit{For the GHZ initial state, the tangle of the outgoing state in the Breit-Wheeler process is
\begin{equation}
\tau_{ABC}=\left(\frac{a_+^2-a_-^2}{a_+^2+a_-^2}\right)^2=\left(\frac{2\xi_e\sqrt{\xi_e^2-1}}{2\xi_e^2-1}\right)^2,
\label{eq:bw-ghz-tangle}
\end{equation}
which is independent of $\theta$ and explicitly saturates the bound of Eq.~\eqref{eq:ghz-bound}.}
\smallskip

\noindent\textit{Proof.} As detailed in Appendix~\ref{app:zero-tangle}, both branch matrices are diagonal, reducing the discriminant to $(a_+^2-a_-^2)^2$. The common $\theta$-dependent factor cancels upon normalisation. \hfill$\square$
\medskip

The amplitudes $a_+$ and $a_-$ represent producing a fermion pair with the same or opposite helicity labels as the incoming photons. At high energy, helicity conservation forces $a_+\gg a_-$, driving $\tau_{ABC}\to1$. Near threshold ($\xi_e\to1^+$), the amplitudes equalise and the branches become parallel. Physically, the pair is produced at rest in an $S$ wave, where $J=1$ is forbidden for two photons by the Landau-Yang theorem. Both branches project onto the single $^1S_0$ state, completely decoupling $C$. 

In both electron-muon and Bhabha scattering, entanglement is suppressed only in a small region near $\theta=\pi$ at low momentum [Figs.~\ref{fig:Tangle_bhabha-initial-ghz} and~\ref{fig:Tangle_electron-muon-initial-ghz}]. The electron (or positron) enters and leaves each branch with that branch's helicity; because it rarely flips, it acts as a robust tag, keeping the branches orthogonal. The branch overlap strictly tracks the flip probability, which is appreciable only for backward scattering at low momentum. Conversely to generation, preserving entanglement requires a particle that cannot easily flip its helicity.

M\o{}ller and Compton scattering preserve entanglement at high momentum but suppress it at low momentum [Figs.~\ref{fig:Tangle_moller-initial-ghz} and~\ref{fig:Tangle_compton-initial-ghz}]. In M\o{}ller scattering, the suppression band centres on $\theta=\pi/2$. In Compton scattering, the band sits at $\theta=\pi/2$ for the lowest momenta, then weakens and drifts backwards as momentum grows. In both M\o{}ller scattering and photon-pair production, this suppression peaks where the first initial state produces the most tangle. The same helicity mixing that creates tripartite entanglement from a product state simultaneously scrambles an existing one.

Photon-pair production uniquely suppresses GHZ entanglement at high momentum [Fig.~\ref{fig:Tangle_photon-pair-production-initial-ghz}]. As derived in Appendix~\ref{app:zero-tangle}, its tangle follows the closed form:
\begin{equation}
\tau_{ABC}=\frac{4\xi_e^2(\xi_e^2+1)\cos^2\theta\,(1+\sin^2\theta)}{\left(2\xi_e^2+1+\xi_e^2\sin^4\theta\right)^2}.
\label{eq:gg-ghz-tangle}
\end{equation}
At high momentum, the tangle becomes independent of $\xi_e$, remaining large only at forward and backward angles. The $\cos^2\theta$ numerator drives the most striking feature: the tangle vanishes entirely at $\theta=\pi/2$ for \textit{all} momenta, even though Eq.~\eqref{eq:ghz-bound} permits a large value (meaning branch overlap is not the cause). However, the GGM and $\pi$-tangle remain positive (Fig.~\ref{fig:ghz-photon-pair}). At $\xi_e=1/\sqrt3$ (where the incoming leptons travel at $0.5c$), the outgoing state is locally unitarily equivalent to the W state (see Appendix~\ref{app:zero-tangle}). The collision actively converts the initial GHZ state into a W-class state.

In muon-pair production, all three measures vanish for the GHZ initial state across all momenta and angles, leading to a kinematic rule:

\medskip
\noindent\textbf{Statement 3.} \textit{In any pure $s$-channel QED annihilation, the spectator factorises from the outgoing state for the GHZ initial state, destroying all tripartite entanglement.}
\smallskip

\noindent\textit{Proof.} Both branches of the GHZ initial state carry $J_z=0$. By Eq.~\eqref{eq:s-channel-factorisation}, both funnel through the identical longitudinal polarisation of the virtual photon, making them proportional to the same state $\ket{\chi_0}$. These parallel branches force $C$ to factor out of Eq.~\eqref{eq:generic-branches}. \hfill$\square$
\medskip

For muon-pair production, $\ket{\phi_{\mathrm{NN}}}=-\ket{\phi_{\mathrm{PP}}}$, collapsing the outgoing state to:
\begin{equation}
\ket{f}\propto\ket{\Phi_\mu}_{AB}\otimes\ket{x_-}_C, \qquad \ket{x_-}=\frac{1}{\sqrt{2}}\left(\ket{\mathrm{P}}-\ket{\mathrm{N}}\right),
\label{eq:mupair-ghz-final}
\end{equation}
leaving the muon pair $\ket{\Phi_\mu}$ maximally entangled and the spectator stranded in the fixed pure state $\ket{x_-}$. Physically, the annihilation acts as a rank-one projective measurement. Only the specific $J_z=0$ combination coupling to the longitudinal polarisation can annihilate; post-selecting on the event erases all memory of the original branch, decoupling the spectator.

\begin{figure*}[t]
\centering
\subfloat[M\o{}ller, uncorrelated $A$ \label{fig:svetlichny-moller-separable}]{%
    \includegraphics[width=0.32\linewidth]{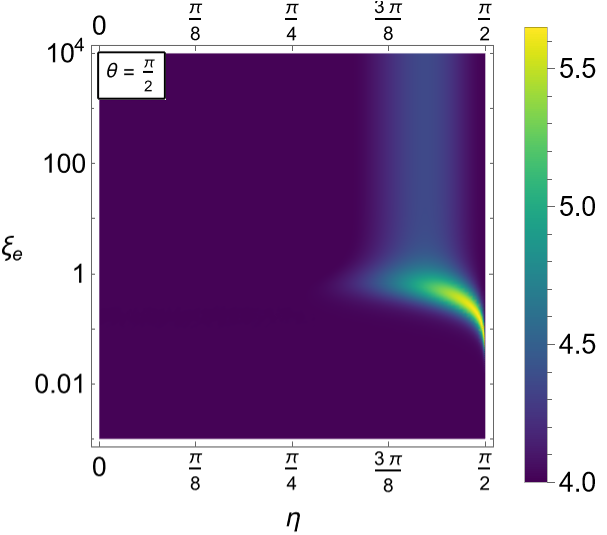}}\hfill
\subfloat[Bhabha, uncorrelated $A$ \label{fig:svetlichny-bhabha-separable}]{%
    \includegraphics[width=0.32\linewidth]{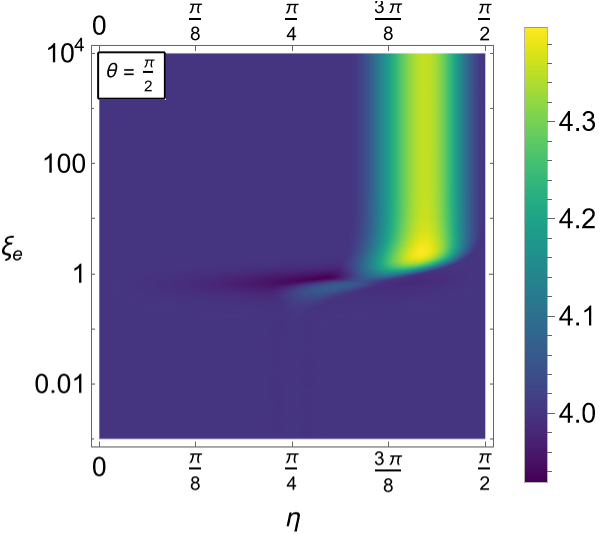}}\hfill
\subfloat[Breit-Wheeler, uncorrelated $A$ \label{fig:svetlichny-breit-wheeler-separable}]{%
    \includegraphics[width=0.32\linewidth]{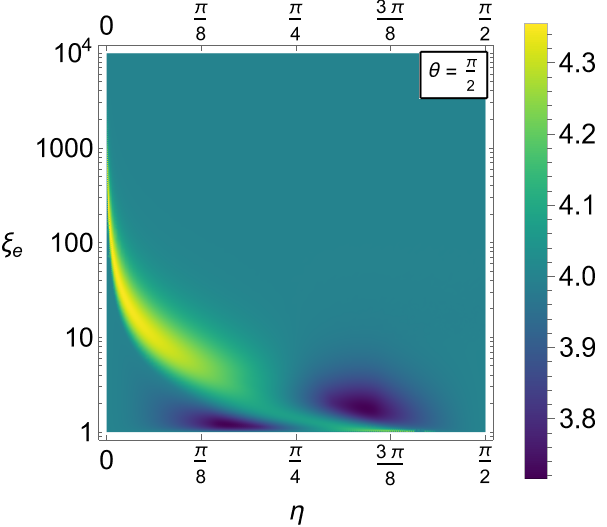}}\\
\subfloat[Photon pair, uncorrelated $A$ \label{fig:svetlichny-photon-pair-separable}]{%
    \includegraphics[width=0.32\linewidth]{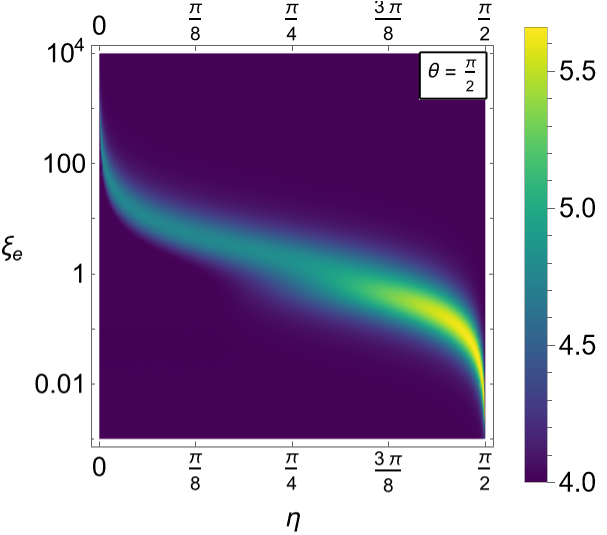}}\hfill
\subfloat[Muon pair, uncorrelated $A$ \label{fig:svetlichny-muon-pair-separable}]{%
    \includegraphics[width=0.32\linewidth]{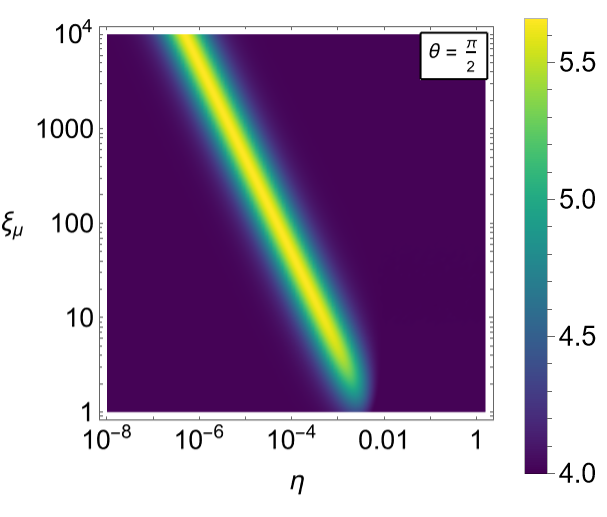}}\hfill
\subfloat[M\o{}ller, GHZ \label{fig:svetlichny-moller-ghz}]{%
    \includegraphics[width=0.32\linewidth]{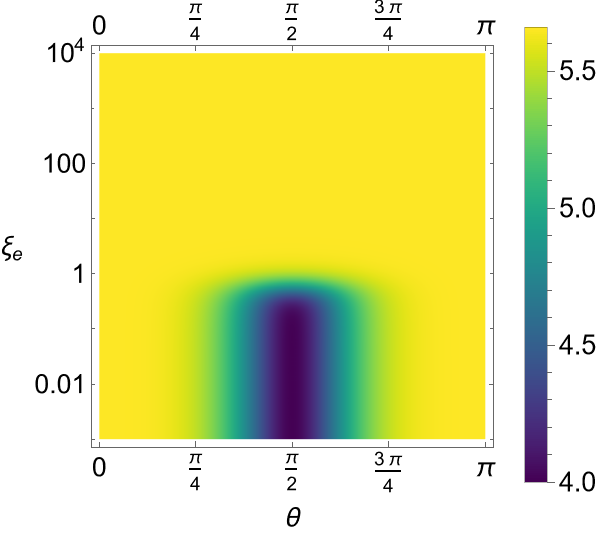}}\\
\subfloat[Bhabha, GHZ \label{fig:svetlichny-bhabha-ghz}]{%
    \includegraphics[width=0.32\linewidth]{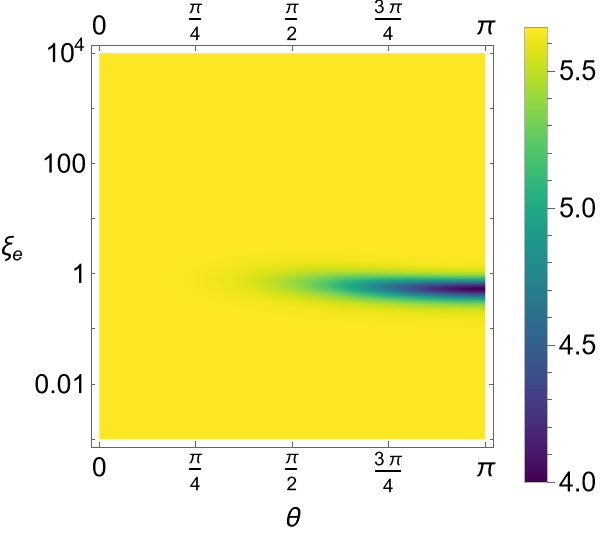}}\hfill
\subfloat[Breit-Wheeler, GHZ \label{fig:svetlichny-breit-wheeler-ghz}]{%
    \includegraphics[width=0.32\linewidth]{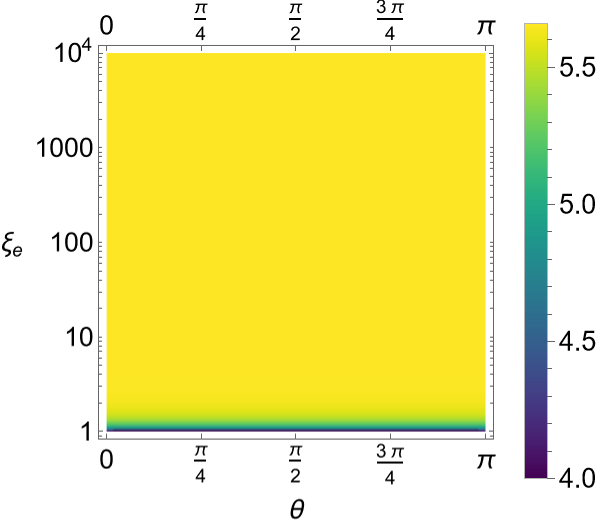}}\hfill
\subfloat[Photon pair, GHZ \label{fig:svetlichny-photon-pair-ghz}]{%
    \includegraphics[width=0.32\linewidth]{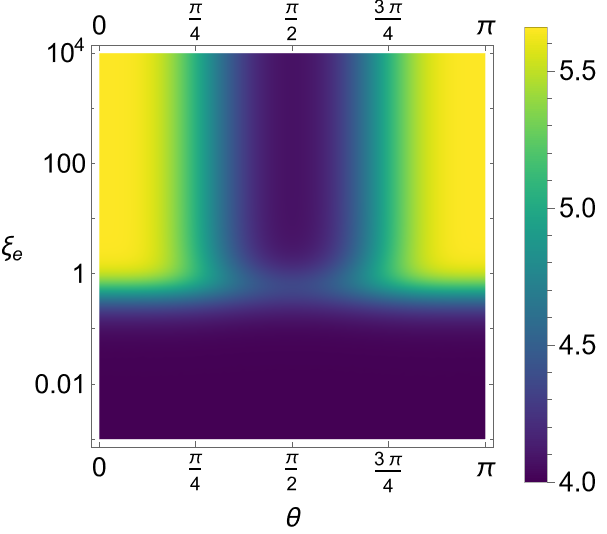}}
\caption{\label{fig:svetlichny-grid} Lower bound on the maximal Svetlichny value, $Q_{\text{Sv}}^{\text{LB}}$ [Eq.~\eqref{eq:svetlichny-seesaw-lb}]. Maps for the uncorrelated-$A$ input, (a)--(e), are at $\theta=\pi/2$ in the $(\eta,\xi)$ plane, with $\xi_\mu$ for muon-pair production and $\xi_e$ otherwise. Maps for the GHZ input, (f)--(i), are in the $(\theta,\xi_e)$ plane.}

\end{figure*}
\begin{table*}[t]
\centering
\caption{\label{tab:svetlichny-points}See-saw lower bound $Q_{\text{Sv}}^{\text{LB}}$ [Eq.~\eqref{eq:svetlichny-seesaw-lb}], numerically maximised Svetlichny value $Q_{\text{Sv}}$, and analytical upper bound $Q_{\text{Sv}}^{\text{UB}}$ [Eq.~\eqref{eq:svetlichny-min-upper-bound}] at representative points. Every point is evaluated at $\theta=\pi/2$. Values with $Q_{\text{Sv}}^{\text{LB}}>4$ certify a violation. As checks, the same optimisation gives $4\sqrt2\approx5.66$ for the GHZ state and $4.35$ for the W state.}
{\renewcommand{\arraystretch}{1.2}\small
\begin{tabular*}{\linewidth}{@{\extracolsep{\fill}}llllll}
\hline\hline
Scattering process & Incoming state & Kinematics & $Q_{\text{Sv}}^{\text{LB}}$ & $Q_{\text{Sv}}$ & $Q_{\text{Sv}}^{\text{UB}}$ \\
\hline
M\o{}ller & Uncorrelated $A$ & $\xi_e=0.5$, $\eta=0.46\pi$ & $5.34$ & $5.34$ & $5.57$ \\
M\o{}ller & Uncorrelated $A$ & $\xi_e=30$, $\eta=0.42\pi$ & $4.35$ & $4.35$ & $5.45$ \\
Photon-pair production & Uncorrelated $A$ & $\xi_e=0.05$, $\eta=0.48\pi$ & $5.58$ & $5.58$ & $5.58$ \\
Photon-pair production & Uncorrelated $A$ & $\xi_e=10$, $\eta=0.07\pi$ & $4.75$ & $4.75$ & $5.39$ \\
Muon-pair production & Uncorrelated $A$ & $\xi_\mu=1$, $\eta=2.4\times10^{-3}$ & $4.35$ & $4.35$ & $5.49$ \\
Muon-pair production & Uncorrelated $A$ & $\xi_\mu=30$, $\eta=1.6\times10^{-4}$ & $5.65$ & $5.65$ & $5.65$ \\
\hline
M\o{}ller & GHZ & $\xi_e=0.05$ & $4.00$ & $4.00$ & $4.00$ \\
M\o{}ller & GHZ & $\xi_e=0.3$ & $4.17$ & $4.17$ & $4.17$ \\
Photon-pair production & GHZ & $\xi_e=1/\sqrt3$ & $4.35$ & $4.35$ & $5.50$ \\
Photon-pair production & GHZ & $\xi_e=10$ & $4.08$ & $4.08$ & $4.43$ \\
Breit-Wheeler & GHZ & $\xi_e=1.02$ & $4.28$ & $4.28$ & $4.28$ \\\hline\hline
\end{tabular*}}
\end{table*}
\begin{figure}[!t]
\centering
\subfloat[Compton scattering \label{fig:svetlichny-compton-ghz}]{%
    \includegraphics[width=0.6\columnwidth]{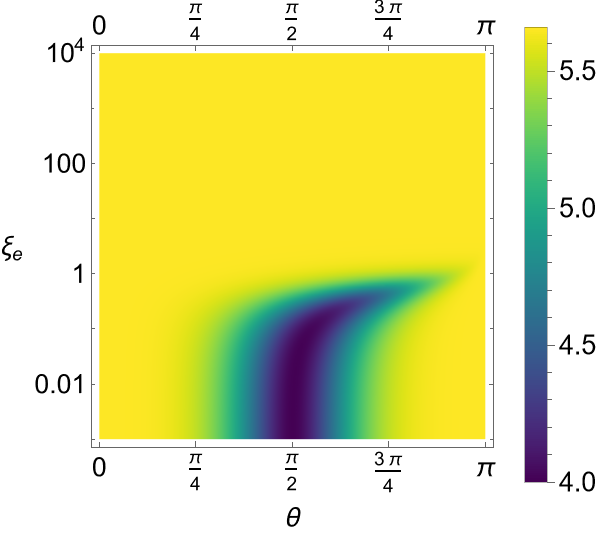}}\\
\subfloat[Electron-muon scattering \label{fig:svetlichny-electron-muon-ghz}]{%
    \includegraphics[width=0.6\columnwidth]{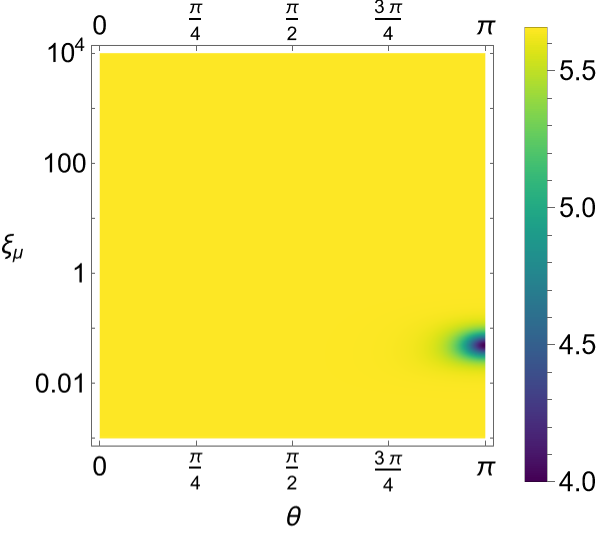}}
\caption{\label{fig:svetlichny-ghz-compton-electron-muon} Lower bound on the maximal Svetlichny value, $Q_{\text{Sv}}^{\text{LB}}$ [Eq.~\eqref{eq:svetlichny-seesaw-lb}], for the GHZ input in (a) Compton and (b) electron-muon scattering, in the $(\theta,\xi)$ plane with $\xi_e$ and $\xi_\mu$ respectively.}
\end{figure}

\section{Genuine tripartite nonlocality} \label{sec:svetlichny-results}
\noindent
To determine whether the outgoing states exhibit genuine tripartite nonlocality, we map the certified lower bound $Q_{\text{Sv}}^{\text{LB}}$ of Eq.~\eqref{eq:svetlichny-seesaw-lb} across the parameter space. Any region with $Q_{\text{Sv}}^{\text{LB}}>4$ certifies a violation of Svetlichny's inequality, whereas values at or below $4$ remain inconclusive.

For the uncorrelated $A$ initial state, Fig.~\ref{fig:svetlichny-grid} shows the resulting maps at $\theta=\pi/2$. M\o{}ller scattering, Bhabha scattering, the Breit-Wheeler process, and photon-pair production exhibit extended bands above $4$ [Figs.~\ref{fig:svetlichny-moller-separable}--\ref{fig:svetlichny-photon-pair-separable}]. These bands closely track the regions of large GGM and $\pi$-tangle identified in Sec.~\ref{sec:tripartite-correlation-results}. In muon-pair production, the violating region narrows to the small-$\eta$ ridge at $\tan\eta\approx m_e/E_e$, where tripartite entanglement first emerges [Fig.~\ref{fig:svetlichny-muon-pair-separable}]. For Compton and electron-muon scattering, $Q_{\text{Sv}}^{\text{LB}}$ exceeds $4$ only over negligible regions of parameter space, so their plots are omitted.

For the GHZ initial state, $Q_{\text{Sv}}^{\text{LB}}$ closely follows the tangle maps of Fig.~\ref{fig:ghz_tangle}, dropping wherever the collision suppresses entanglement [Figs.~\ref{fig:svetlichny-moller-ghz}--\ref{fig:svetlichny-photon-pair-ghz} and~\ref{fig:svetlichny-ghz-compton-electron-muon}]. Consistent with Statement~3, the spectator factorises in muon-pair production, yielding no genuine tripartite nonlocality.

Table~\ref{tab:svetlichny-points} shows the independently maximised Svetlichny value $Q_{\text{Sv}}$ at representative points along with the see-saw lower bound $Q_{\text{Sv}}^{\text{LB}}$ and the analytical upper bound $Q_{\text{Sv}}^{\text{UB}}$ of Eq.~\eqref{eq:svetlichny-min-upper-bound}. From there, we see that $Q_{\text{Sv}}^{\text{LB}}$ matches $Q_{\text{Sv}}$ at every tested point, confirming that the see-saw search reliably finds the maximum irrespective of whether the upper bound is loose or not. For the GHZ input, $Q_{\text{Sv}}^{\text{LB}}$ and $Q_{\text{Sv}}^{\text{UB}}$ meet across the entire parameter space in every process except photon-pair production; they also meet along the near-maximal GHZ ridges of photon-pair and muon-pair production from an uncorrelated $A$. Away from these GHZ-like regimes, a clear gap opens between the two bounds. This happens both for W-class states, such as photon-pair production from a GHZ input at $\theta=\pi/2$ and $\xi_e=1/\sqrt{3}$ (where $Q_{\text{Sv}}^{\text{LB}}$ stays at the W-state limit of $4.35$ while $Q_{\text{Sv}}^{\text{UB}}$ reaches $5.50$), and for generic GHZ-class states, such as uncorrelated-$A$ photon-pair production at $\xi_e=10$ or M\o{}ller scattering at $\xi_e=0.5$. Mapping $Q_{\text{Sv}}^{\text{LB}}$ directly is therefore necessary, as the analytical upper bound overestimates $Q_{\text{Sv}}$ in these regimes and hides features such as the $4.35$ ceiling in W-dominated regions.

\begin{table*}
\centering
\newcommand{\cell}[2]{\parbox[t]{#1}{\raggedright #2\vspace{2pt}}}
\caption{\label{tab:conclusion-summary} Summary of the results. The second column refers to the input with $A$ uncorrelated and $B$ and $C$ partially entangled, the third to the GHZ input, and the last to the mechanism discussed in Sec.~\ref{sec:results}.}
{\renewcommand{\arraystretch}{1.2}\small
\begin{tabular*}{\linewidth}{@{\extracolsep{\fill}}llll}
\hline\hline
\cell{0.14\textwidth}{Scattering process} & \cell{0.26\textwidth}{Generation (uncorrelated $A$)} & \cell{0.26\textwidth}{Survival (GHZ input)} & \cell{0.26\textwidth}{Mechanism} \\
\hline
\cell{0.14\textwidth}{M\o{}ller} & \cell{0.26\textwidth}{GHZ class. Ridge close to $\eta=\pi/2$ at low $\xi_e$, peaked at $\theta=\pi/2$. GGM and $\pi$-tangle persist at high $\xi_e$.} & \cell{0.26\textwidth}{Preserved at high $\xi_e$, suppressed near $\theta=\pi/2$ at low $\xi_e$.} & \cell{0.26\textwidth}{The $t$ and $u$ channels cancel in the ${\mathrm{P}}{\mathrm{N}}$ branch at $\theta=\pi/2$.} \\
\cell{0.14\textwidth}{Bhabha} & \cell{0.26\textwidth}{GHZ class in a narrow band near $\xi_e\approx1$. GGM and $\pi$-tangle persist at high $\xi_e$.} & \cell{0.26\textwidth}{Preserved except in a pocket at low $\xi_e$ and backward angles.} & \cell{0.26\textwidth}{$s$- and $t$-channel interplay.} \\
\cell{0.14\textwidth}{Electron-muon} & \cell{0.26\textwidth}{GHZ class at $\xi_\mu$ of order one, retreating to $\theta\to\pi$, $\eta\to\pi/2$ at higher $\xi_\mu$.} & \cell{0.26\textwidth}{Preserved except in a small spot near $\theta=\pi$ at low $\xi_\mu$.} & \cell{0.26\textwidth}{Helicity flip of the slow muon generates. The electron's rigid helicity preserves.} \\
\cell{0.14\textwidth}{Muon-pair production} & \cell{0.26\textwidth}{GHZ class. Narrow ridge at $\tan\eta\approx m_e/E_e$, at reduced rate.} & \cell{0.26\textwidth}{Zero. The spectator factorises.} & \cell{0.26\textwidth}{Single virtual photon. Suppression for the uncorrelated input, alignment for GHZ.} \\
\cell{0.14\textwidth}{Compton} & \cell{0.26\textwidth}{W class only, zero tangle.} & \cell{0.26\textwidth}{Preserved at high $\xi_e$, suppressed in a central band at low $\xi_e$.} & \cell{0.26\textwidth}{Cancellation in the hyperdeterminant.} \\
\cell{0.14\textwidth}{Photon-pair production} & \cell{0.26\textwidth}{GHZ class. Ridge close to $\eta=\pi/2$ at low $\xi_e$, peaked at $\theta=\pi/2$.} & \cell{0.26\textwidth}{Suppressed at central angles at all $\xi_e$ and at all angles at low $\xi_e$. W class at $\theta=\pi/2$.} & \cell{0.26\textwidth}{Landau-Yang theorem suppresses the ${\mathrm{P}}{\mathrm{N}}$ branch.} \\
\cell{0.14\textwidth}{Breit-Wheeler} & \cell{0.26\textwidth}{W class only, zero tangle.} & \cell{0.26\textwidth}{Preserved except at threshold, independent of $\theta$.} & \cell{0.26\textwidth}{Landau-Yang theorem projects both branches onto $^1S_0$ at threshold.} \\\hline\hline
\end{tabular*}}
\end{table*}
\section{Summary and outlook} \label{sec:conclusion}
\noindent
We investigated how QED scattering generates and preserves genuine multipartite entanglement using a non-interacting spectator. Because the spectator $C$ never interacts, its final correlations arise entirely through its initial entanglement with the colliding particles. We mapped the GME of the outgoing three-qubit helicity state across seven tree-level processes for two initial states: an uncorrelated $A$ and a tripartite GHZ state.

Our findings, summarised in Table~\ref{tab:conclusion-summary}, demonstrate how the nature of a QED process affects entanglement. The spectator serves as a record-keeper of the incoming helicity configuration, splitting the outgoing state into two branches. Tripartite entanglement vanishes whenever the collision suppresses one branch or aligns the two. Because helicity flips provide the primary mechanism to avoid this and cost an amplitude factor of $\mathcal{O}(m/E)$, GME generation peaks at low momentum. Conversely, preserving an initial GHZ state requires avoiding helicity flips, meaning particles like the electron act as robust informational tags at high momentum. As electron-muon scattering demonstrates, the kinematic requirements for generating and preserving entanglement are strictly inverted. Furthermore, because these states are explicitly conditioned on a scattering event, large GME values result from entanglement concentration; the necessary branch imbalance exacts a high cost in event rate, which any resource application must account for.

The scattering processes dictate not only the magnitude but the exact class of the entanglement. For an uncorrelated $A$ initial state, photon-pair production, muon-pair production, M\o{}ller, Bhabha, and electron-muon scattering generate GHZ-class entanglement. By contrast, Compton scattering and the Breit-Wheeler process generate strictly W-class states with an identically vanishing tangle (Statement~1). A collision can also actively convert entanglement classes, as seen when photon-pair production turns a GHZ initial state into a W-class state at $\theta=\pi/2$. For $s$-channel annihilations, the kinematic constraints act as a rank-one projective measurement that completely decouples the spectator from a GHZ initial state (Statement~3), whereas the Breit-Wheeler process (Statement~2), Bhabha scattering, and electron-muon scattering preserve the GHZ state over almost all of phase space.

Beyond entanglement, the outgoing states also exhibit genuine tripartite nonlocality. Mapping a certified see-saw lower bound across parameter space -- and benchmarking it against an analytical upper bound at representative points -- certifies violations of Svetlichny's inequality for both initial states.

Several theoretical questions remain open. Foremost is the algebraic origin of the exact zeros of the tangle in Statement~1 and in photon-pair production at $\theta=\pi/2$. Because the Breit-Wheeler process and photon-pair production are time-reversals of each other, they share the exact same fundamental QED symmetries. However, their tangles behave differently: the Breit-Wheeler tangle vanishes globally for the uncorrelated-$A$ input, whereas the photon-pair tangle vanishes only at $\theta=\pi/2$ for the GHZ input. This demonstrates that these zeros are structural properties of the directional quantum information map rather than simple consequences of the Feynman vertices. Identifying the exact algebraic mechanisms that restrict these specific projective views exclusively to the W class remains an open problem. Finally, future work could incorporate electroweak interactions, where chiral couplings modify the helicity structure, or extend the spectator framework to noisy initial states to study the interplay between QED post-selection and decoherence.

\bibliography{apssamp}

\appendix
\section{Tree-level QED helicity amplitudes} \label{app:qed-helicity-amplitudes}
\noindent
We show the Feynman amplitudes of the seven processes, our spinor and polarisation conventions, and the helicity amplitudes that enter Eq.~\eqref{eq:phi-branches}. We abbreviate $u_i\equiv u(\momentum{p}_i,s_i)$, $v_i\equiv v(\momentum{p}_i,s_i)$ and $\epsilon_{i\mu}\equiv\epsilon_\mu(\momentum{p}_i,s_i)$ for the Dirac spinors of particles and antiparticles and the photon polarisation vectors, and write $\slashed{p}=\gamma^\mu p_\mu$. For processes with external photons, the Dirac equation has been used to simplify the numerators. M\o{}ller scattering receives $t$- and $u$-channel contributions with a relative sign fixed by Fermi statistics,
\begin{align*}
\mathcal{M}_{\text{M}} = e^2\bigg[&\frac{(\bar u_3\gamma^\mu u_1)(\bar u_4\gamma_\mu u_2)}{(p_1-p_3)^2}
- \frac{(\bar u_4\gamma^\mu u_1)(\bar u_3\gamma_\mu u_2)}{(p_1-p_4)^2}\bigg]. 
\end{align*}
Bhabha scattering combines $s$-channel annihilation with $t$-channel exchange,
\begin{align*}
\mathcal{M}_{\text{B}} = e^2\bigg[&\frac{(\bar v_2\gamma^\mu u_1)(\bar u_3\gamma_\mu v_4)}{(p_1+p_2)^2}
- \frac{(\bar u_3\gamma^\mu u_1)(\bar v_2\gamma_\mu v_4)}{(p_1-p_3)^2}\bigg]. 
\end{align*}
Electron-muon scattering and muon-pair production each have a single diagram,
\begin{align*}
\mathcal{M}_{e\mu} &= e^2\,\frac{(\bar u_3\gamma^\mu u_1)(\bar u_4\gamma_\mu u_2)}{(p_1-p_3)^2}, 
\\
\mathcal{M}_{\mu\mu} &= e^2\,\frac{(\bar v_2\gamma^\mu u_1)(\bar u_3\gamma_\mu v_4)}{(p_1+p_2)^2}. 
\end{align*}
The three processes with external photons are
\begin{align*}
\mathcal{M}_{\text{C}} ={}& -e^2\,\epsilon^*_{3\mu}\epsilon_{1\nu}\,\bar u_4\bigg[\frac{\gamma^\mu\slashed{p}_1\gamma^\nu+2p_2^\nu\gamma^\mu}{2p_1\cdot p_2}\\
&\quad+ \frac{2p_2^\mu\gamma^\nu-\gamma^\nu\slashed{p}_3\gamma^\mu}{-2p_2\cdot p_3}\bigg]u_2, \\
\mathcal{M}_{\gamma\gamma} ={}& -e^2\,\epsilon^*_{4\mu}\epsilon^*_{3\nu}\,\bar v_2\bigg[\frac{2p_1^\nu\gamma^\mu-\gamma^\mu\slashed{p}_3\gamma^\nu}{-2p_1\cdot p_3}\\
&\quad+ \frac{2p_1^\mu\gamma^\nu-\gamma^\nu\slashed{p}_4\gamma^\mu}{-2p_1\cdot p_4}\bigg]u_1, \\
\mathcal{M}_{\text{BW}} ={}& -e^2\,\epsilon_{1\mu}\epsilon_{2\nu}\,\bar u_3\bigg[\frac{2p_3^\mu\gamma^\nu-\gamma^\mu\slashed{p}_1\gamma^\nu}{-2p_1\cdot p_3}\\
&\quad+ \frac{2p_3^\nu\gamma^\mu-\gamma^\nu\slashed{p}_2\gamma^\mu}{-2p_2\cdot p_3}\bigg]v_4, 
\end{align*}
for Compton scattering, photon-pair production and the Breit-Wheeler process.

We evaluate these with helicity spinors. For a particle with four-momentum $p=(\omega,|\momentum{p}|\sin\theta\cos\phi,|\momentum{p}|\sin\theta\sin\phi,|\momentum{p}|\cos\theta)$, we take the spinors below, where $p$ inside the square roots stands for $|\momentum{p}|$. Introducing a sign parameter $\zeta = 1$ for the $u$ spinors and $\zeta = -1$ for the $v$ spinors, and a second parameter $\kappa = \pm 1$ to capture the $\momentum{p} \to -\momentum{p}$ transformation, the spinors can be compactly written as,
\begin{align*}
\left. \begin{array}{r}
u_{\mathrm{P}}(\momentum{p}),\, v_{\mathrm{N}}(\momentum{p}) \;\; (\kappa=1) \\
u_{\mathrm{N}}(-\momentum{p}),\, v_{\mathrm{P}}(-\momentum{p}) \;\; (\kappa=-1)
\end{array} \right\} &= 
\begin{pmatrix}
\sqrt{\omega - \kappa p} \cos(\theta/2) \\
 e^{i\phi} \sqrt{\omega - \kappa p} \sin(\theta/2) \\
\zeta\sqrt{\omega + \kappa p} \cos(\theta/2) \\
\zeta e^{i\phi} \sqrt{\omega + \kappa p} \sin(\theta/2)
\end{pmatrix}, \\[2ex]
\left. \begin{array}{r}
u_{\mathrm{N}}(\momentum{p}),\, v_{\mathrm{P}}(\momentum{p}) \;\; (\kappa=1) \\
u_{\mathrm{P}}(-\momentum{p}),\, v_{\mathrm{N}}(-\momentum{p}) \;\; (\kappa=-1)
\end{array} \right\} &= 
\begin{pmatrix}
-\sqrt{\omega + \kappa p} \sin(\theta/2) \\
 e^{i\phi} \sqrt{\omega + \kappa p} \cos(\theta/2) \\
-\zeta\sqrt{\omega - \kappa p} \sin(\theta/2) \\
\zeta e^{i\phi} \sqrt{\omega - \kappa p} \cos(\theta/2)
\end{pmatrix}.
\end{align*}
The circular polarisation vectors of the photon are
\begin{align*}
\epsilon_{{\mathrm{P}}}(\theta,\phi) = \frac{e^{i\phi}}{\sqrt{2}}&\big(0,-\cos\theta\cos\phi+i\sin\phi,\\
&\quad -\cos\theta\sin\phi-i\cos\phi,\sin\theta\big), \\
\epsilon_{{\mathrm{N}}}(\theta,\phi) = \frac{e^{-i\phi}}{\sqrt{2}}&\big(0,\cos\theta\cos\phi+i\sin\phi,\\
&\quad \cos\theta\sin\phi-i\cos\phi,-\sin\theta\big).
\end{align*}

All amplitudes are evaluated in the COM frame of $A$ and $B$, with $p_1, p_2$ the incoming and $p_3, p_4$ the outgoing four-momenta. We set the scattering plane to $\phi=0$, placing the incoming particles along the $z$-axis and the outgoing ones at the scattering angle $\theta$:
\begin{align*}
    p_1 &=(\omega_1,0,0,|\momentum{p}_1|), \\
    p_2 &=(\omega_2,0,0,-|\momentum{p}_2|),\\
    p_3 &=(\omega_3,|\momentum{p}_3|\sin\theta,0,|\momentum{p}_3|\cos\theta), \\
    p_4 &=(\omega_4,-|\momentum{p}_4|\sin\theta,0,-|\momentum{p}_4|\cos\theta),
\end{align*}
with $|\momentum{p}_1|=|\momentum{p}_2|=|\momentum{p}|$. Inserting the spinors and polarisation vectors into the Feynman amplitudes gives the helicity amplitudes below, written in terms of $\xi_e$, $\xi_\mu$ and $\lambda_m$. The common factor $e^2$ is dropped, since it cancels upon normalisation.

\subsection{Photon-pair production}
\noindent
For $e^-e^+\rightarrow\gamma\gamma$, we have $\omega = \sqrt{m^2_e+|\momentum{p}|^2}$ and outgoing momenta $|\momentum{p}_3|=|\momentum{p}_4|=\omega$. Defining $\Xi_e^+ = \sqrt{\xi_e^2 + 1}$ and the denominator $\Delta = \xi_e^2 \sin^2\theta + 1$, the non-zero amplitudes are:
\begin{align*}
\mathcal{M}_{{\mathrm{P}}{\mathrm{P}};{\mathrm{P}}{\mathrm{P}}} &=-\mathcal{M}_{{\mathrm{N}}{\mathrm{N}};{\mathrm{N}}{\mathrm{N}}} = -\frac{2(\xi_e+\Xi_e^+)}{\Delta}, \\
\mathcal{M}_{{\mathrm{P}}{\mathrm{P}};{\mathrm{N}}{\mathrm{N}}} &=-\mathcal{M}_{{\mathrm{N}}{\mathrm{N}};{\mathrm{P}}{\mathrm{P}}} = \frac{2(\Xi_e^+-\xi_e)}{\Delta}, \\
\mathcal{M}_{{\mathrm{P}}{\mathrm{P}};{\mathrm{P}}{\mathrm{N}}} &=\mathcal{M}_{{\mathrm{P}}{\mathrm{P}};{\mathrm{N}}{\mathrm{P}}} = -\mathcal{M}_{{\mathrm{N}}{\mathrm{N}};{\mathrm{P}}{\mathrm{N}}} = -\mathcal{M}_{{\mathrm{N}}{\mathrm{N}};{\mathrm{N}}{\mathrm{P}}} = \frac{2 \xi_e \sin^2\theta}{\Delta}, \\
\mathcal{M}_{{\mathrm{P}}{\mathrm{N}};{\mathrm{P}}{\mathrm{N}}} &=\mathcal{M}_{{\mathrm{N}}{\mathrm{P}};{\mathrm{N}}{\mathrm{P}}} = -\frac{2 \xi_e \Xi_e^+ \sin\theta (\cos\theta+1)}{\Delta}, \\
\mathcal{M}_{{\mathrm{P}}{\mathrm{N}};{\mathrm{N}}{\mathrm{P}}} &=\mathcal{M}_{{\mathrm{N}}{\mathrm{P}};{\mathrm{P}}{\mathrm{N}}} = \frac{2 \xi_e \Xi_e^+ \sin\theta (1-\cos\theta)}{\Delta}.
\end{align*}
The remaining amplitudes (from incoming ${\mathrm{P}}{\mathrm{N}}$ or ${\mathrm{N}}{\mathrm{P}}$ to outgoing ${\mathrm{P}}{\mathrm{P}}$ or ${\mathrm{N}}{\mathrm{N}}$) vanish.

\subsection{Breit-Wheeler process}
\noindent
For $\gamma\gamma\rightarrow e^-e^+$, we have $\omega = |\momentum{p}|$ and outgoing momenta $|\momentum{p}_3|=|\momentum{p}_4| = \sqrt{|\momentum{p}|^2 - m^2_e}$. Defining $\Xi_e^- = \sqrt{\xi_e^2 - 1}$ and the denominator $\Delta = (\xi_e^2-1) \cos^2\theta - \xi_e^2$, the non-zero amplitudes are:
\begin{align*}
\mathcal{M}_{{\mathrm{P}}{\mathrm{P}};{\mathrm{P}}{\mathrm{P}}} &=-\mathcal{M}_{{\mathrm{N}}{\mathrm{N}};{\mathrm{N}}{\mathrm{N}}} = \frac{2(\xi_e+\Xi_e^-)}{\Delta}, \\
\mathcal{M}_{{\mathrm{P}}{\mathrm{P}};{\mathrm{N}}{\mathrm{N}}} &=-\mathcal{M}_{{\mathrm{N}}{\mathrm{N}};{\mathrm{P}}{\mathrm{P}}} = \frac{2(\xi_e-\Xi_e^-)}{\Delta}, \\
\mathcal{M}_{{\mathrm{P}}{\mathrm{N}};{\mathrm{P}}{\mathrm{P}}} &=-\mathcal{M}_{{\mathrm{P}}{\mathrm{N}};{\mathrm{N}}{\mathrm{N}}} = -\mathcal{M}_{{\mathrm{N}}{\mathrm{P}};{\mathrm{N}}{\mathrm{N}}} = \mathcal{M}_{{\mathrm{N}}{\mathrm{P}};{\mathrm{P}}{\mathrm{P}}} = -\frac{2 \Xi_e^- \sin^2\theta}{\Delta}, \\
\mathcal{M}_{{\mathrm{P}}{\mathrm{N}};{\mathrm{P}}{\mathrm{N}}} &=\mathcal{M}_{{\mathrm{N}}{\mathrm{P}};{\mathrm{N}}{\mathrm{P}}} = \frac{2 \xi_e \Xi_e^- \sin\theta (1+\cos\theta)}{\Delta}, \\
\mathcal{M}_{{\mathrm{P}}{\mathrm{N}};{\mathrm{N}}{\mathrm{P}}} &=\mathcal{M}_{{\mathrm{N}}{\mathrm{P}};{\mathrm{P}}{\mathrm{N}}} = -\frac{2 \xi_e \Xi_e^- \sin\theta (1-\cos\theta)}{\Delta}.
\end{align*}
The remaining amplitudes (from incoming ${\mathrm{P}}{\mathrm{P}}$ or ${\mathrm{N}}{\mathrm{N}}$ to outgoing ${\mathrm{P}}{\mathrm{N}}$ or ${\mathrm{N}}{\mathrm{P}}$) vanish.

\subsection{Compton scattering}
\noindent
For $\gamma e^-\rightarrow\gamma e^-$, the incoming and outgoing photons have momenta $|\momentum{p}_1|=|\momentum{p}_3|=|\momentum{p}|$ and the electrons have energy $\omega_e = \sqrt{|\momentum{p}|^2+m_e^2}$. Defining $\Xi_e^+ = \sqrt{\xi_e^2+1}$ and the denominator $\Delta = \xi_e \cos\theta + \Xi_e^+$, the amplitudes are:
\begin{align*}
\mathcal{M}_{{\mathrm{P}}{\mathrm{P}};{\mathrm{P}}{\mathrm{P}}} &=\mathcal{M}_{{\mathrm{N}}{\mathrm{N}};{\mathrm{N}}{\mathrm{N}}} = -\frac{\cos(\theta/2) \left[4 \xi_e(\xi_e+\Xi_e^+) + \cos\theta + 1\right]}{(\xi_e+\Xi_e^+)\Delta}, \\
\mathcal{M}_{{\mathrm{P}}{\mathrm{P}};{\mathrm{P}}{\mathrm{N}}} &=-\mathcal{M}_{{\mathrm{N}}{\mathrm{N}};{\mathrm{N}}{\mathrm{P}}} = -\frac{2 \sin(\theta/2) \cos^2(\theta/2) (\Xi_e^+ - \xi_e \cos\theta)}{\xi_e^2 \sin^2\theta + 1}, \\
\mathcal{M}_{{\mathrm{P}}{\mathrm{P}};{\mathrm{N}}{\mathrm{P}}} &=\mathcal{M}_{{\mathrm{N}}{\mathrm{N}};{\mathrm{P}}{\mathrm{N}}} = -\frac{\sin(\theta/2) \sin\theta}{(\xi_e+\Xi_e^+)\Delta}, \\
\mathcal{M}_{{\mathrm{P}}{\mathrm{P}};{\mathrm{N}}{\mathrm{N}}} &=-\mathcal{M}_{{\mathrm{N}}{\mathrm{N}};{\mathrm{P}}{\mathrm{P}}} = \frac{2 (\xi_e-\Xi_e^+) \sin^3(\theta/2)}{(\xi_e+\Xi_e^+)\Delta}, \\
\mathcal{M}_{{\mathrm{P}}{\mathrm{N}};{\mathrm{P}}{\mathrm{P}}} &=-\mathcal{M}_{{\mathrm{N}}{\mathrm{P}};{\mathrm{N}}{\mathrm{N}}} = \frac{\sin^2\theta \csc(\theta/2)}{2\Delta}, \\
\mathcal{M}_{{\mathrm{P}}{\mathrm{N}};{\mathrm{N}}{\mathrm{N}}} &=\mathcal{M}_{{\mathrm{N}}{\mathrm{P}};{\mathrm{P}}{\mathrm{P}}} = \frac{(\xi_e-\Xi_e^+) \sin(\theta/2) \sin\theta}{\Delta}, \\
\mathcal{M}_{{\mathrm{P}}{\mathrm{N}};{\mathrm{P}}{\mathrm{N}}} &=\mathcal{M}_{{\mathrm{N}}{\mathrm{P}};{\mathrm{N}}{\mathrm{P}}} = -\frac{2 (\xi_e+\Xi_e^+) \cos^3(\theta/2)}{\Delta}, \\
\mathcal{M}_{{\mathrm{P}}{\mathrm{N}};{\mathrm{N}}{\mathrm{P}}} &=-\mathcal{M}_{{\mathrm{N}}{\mathrm{P}};{\mathrm{P}}{\mathrm{N}}} = \frac{2 \sin^3(\theta/2)}{\Delta}.
\end{align*}

\subsection{M\o{}ller scattering}
\noindent
For $e^-e^-\rightarrow e^-e^-$, all particles have energy $\omega = \sqrt{|\momentum{p}|^2+m_e^2}$ and momentum magnitude $|\momentum{p}|$. Defining $\Xi_e^+ = \sqrt{\xi_e^2+1}$, the amplitudes are:
\begin{align*}
\mathcal{M}_{{\mathrm{P}}{\mathrm{P}};{\mathrm{P}}{\mathrm{P}}} &=\mathcal{M}_{{\mathrm{N}}{\mathrm{N}};{\mathrm{N}}{\mathrm{N}}} = \frac{2-4(2\xi_e^2+1)\csc^2\theta}{\xi_e^2}, \\
\mathcal{M}_{{\mathrm{P}}{\mathrm{P}};{\mathrm{N}}{\mathrm{N}}} &=\mathcal{M}_{{\mathrm{N}}{\mathrm{N}};{\mathrm{P}}{\mathrm{P}}} = \frac{2}{\xi_e^2}, \\
\mathcal{M}_{{\mathrm{P}}{\mathrm{P}};{\mathrm{N}}{\mathrm{P}}} &=\mathcal{M}_{{\mathrm{N}}{\mathrm{N}};{\mathrm{N}}{\mathrm{P}}} = \mathcal{M}_{{\mathrm{P}}{\mathrm{N}};{\mathrm{P}}{\mathrm{P}}} = \mathcal{M}_{{\mathrm{P}}{\mathrm{N}};{\mathrm{N}}{\mathrm{N}}} = \frac{2\Xi_e^+ \cot\theta}{\xi_e^2}, \\
\mathcal{M}_{{\mathrm{P}}{\mathrm{P}};{\mathrm{P}}{\mathrm{N}}} &=\mathcal{M}_{{\mathrm{N}}{\mathrm{N}};{\mathrm{P}}{\mathrm{N}}} = \mathcal{M}_{{\mathrm{N}}{\mathrm{P}};{\mathrm{P}}{\mathrm{P}}} = \mathcal{M}_{{\mathrm{N}}{\mathrm{P}};{\mathrm{N}}{\mathrm{N}}} = -\frac{2\Xi_e^+ \cot\theta}{\xi_e^2}, \\
\mathcal{M}_{{\mathrm{P}}{\mathrm{N}};{\mathrm{P}}{\mathrm{N}}} &=\mathcal{M}_{{\mathrm{N}}{\mathrm{P}};{\mathrm{N}}{\mathrm{P}}} = \frac{2\left[(\xi_e^2+1)\cos\theta+\xi_e^2\right]}{\xi_e^2(\cos\theta-1)}, \\
\mathcal{M}_{{\mathrm{P}}{\mathrm{N}};{\mathrm{N}}{\mathrm{P}}} &=\mathcal{M}_{{\mathrm{N}}{\mathrm{P}};{\mathrm{P}}{\mathrm{N}}} = \frac{2\xi_e^2\tan^2(\theta/2)+\sec^2(\theta/2)-2}{\xi_e^2}.
\end{align*}

\subsection{Bhabha scattering}
\noindent
For $e^-e^+\rightarrow e^-e^+$, all particles have energy $\omega = \sqrt{|\momentum{p}|^2+m_e^2}$ and momentum magnitude $|\momentum{p}|$. Defining $\Xi_e^+ = \sqrt{\xi_e^2+1}$, the amplitudes are:
\begin{align*}
\mathcal{M}_{{\mathrm{P}}{\mathrm{P}};{\mathrm{P}}{\mathrm{P}}} &=\mathcal{M}_{{\mathrm{N}}{\mathrm{N}};{\mathrm{N}}{\mathrm{N}}}\\ &= \frac{\csc^2(\theta/2) \left[\xi_e^2\cos^2\theta + 4\xi_e^4 + 5\xi_e^2 + \cos\theta + 1\right]}{2\xi_e^2(\xi_e^2+1)}, \\
\mathcal{M}_{{\mathrm{P}}{\mathrm{P}};{\mathrm{N}}{\mathrm{N}}} &=\mathcal{M}_{{\mathrm{N}}{\mathrm{N}};{\mathrm{P}}{\mathrm{P}}} = \frac{2\xi_e^2\cos^2(\theta/2)+1}{\xi_e^2(\xi_e^2+1)}, \\
\mathcal{M}_{{\mathrm{P}}{\mathrm{P}};{\mathrm{P}}{\mathrm{N}}} &=\mathcal{M}_{{\mathrm{P}}{\mathrm{P}};{\mathrm{N}}{\mathrm{P}}} = -\mathcal{M}_{{\mathrm{N}}{\mathrm{N}};{\mathrm{P}}{\mathrm{N}}} = -\mathcal{M}_{{\mathrm{N}}{\mathrm{N}};{\mathrm{N}}{\mathrm{P}}} \\
&= -\frac{\cot(\theta/2)(\xi_e^2\cos\theta+1)}{\xi_e^2\Xi_e^+}, \\
\mathcal{M}_{{\mathrm{P}}{\mathrm{N}};{\mathrm{P}}{\mathrm{P}}} &=\mathcal{M}_{{\mathrm{N}}{\mathrm{P}};{\mathrm{P}}{\mathrm{P}}} = -\mathcal{M}_{{\mathrm{P}}{\mathrm{N}};{\mathrm{N}}{\mathrm{N}}} = -\mathcal{M}_{{\mathrm{N}}{\mathrm{P}};{\mathrm{N}}{\mathrm{N}}} \\
&= \frac{\cot(\theta/2)(\xi_e^2\cos\theta+1)}{\xi_e^2\Xi_e^+}, \\
\mathcal{M}_{{\mathrm{P}}{\mathrm{N}};{\mathrm{P}}{\mathrm{N}}} &=\mathcal{M}_{{\mathrm{N}}{\mathrm{P}};{\mathrm{N}}{\mathrm{P}}} = \frac{\cot^2(\theta/2)\left[\xi_e^2(\cos\theta+1)+1\right]}{\xi_e^2}, \\
\mathcal{M}_{{\mathrm{P}}{\mathrm{N}};{\mathrm{N}}{\mathrm{P}}} &=\mathcal{M}_{{\mathrm{N}}{\mathrm{P}};{\mathrm{P}}{\mathrm{N}}} = \frac{\xi_e^2(1-\cos\theta)-1}{\xi_e^2}.
\end{align*}

\subsection{Electron-muon scattering}
\noindent
For $\mu^-e^-\rightarrow\mu^-e^-$, the muons have energy $\omega_\mu = \sqrt{|\momentum{p}|^2+m_\mu^2}$ and the electrons have energy $\omega_e = \sqrt{|\momentum{p}|^2+m_e^2}$. Defining $\Xi_\mu^+ = \sqrt{\xi_\mu^2+1}$ and $\Lambda_\mu = \sqrt{\xi_\mu^2+\lambda_m^2}$, the amplitudes are:
\begin{align*}
\mathcal{M}_{{\mathrm{P}}{\mathrm{P}};{\mathrm{P}}{\mathrm{P}}} &=\mathcal{M}_{{\mathrm{N}}{\mathrm{N}};{\mathrm{N}}{\mathrm{N}}} = \frac{\cot^2(\theta/2) (\Xi_\mu^+ \Lambda_\mu + \xi_\mu^2) + 2\xi_\mu^2}{\xi_\mu^2}, \\
\mathcal{M}_{{\mathrm{P}}{\mathrm{N}};{\mathrm{P}}{\mathrm{N}}} &=\mathcal{M}_{{\mathrm{N}}{\mathrm{P}};{\mathrm{N}}{\mathrm{P}}} = \frac{\cot^2(\theta/2) (\Xi_\mu^+ \Lambda_\mu + \xi_\mu^2)}{\xi_\mu^2}, \\
\mathcal{M}_{{\mathrm{P}}{\mathrm{P}};{\mathrm{P}}{\mathrm{N}}} &=-\mathcal{M}_{{\mathrm{N}}{\mathrm{N}};{\mathrm{N}}{\mathrm{P}}} = -\mathcal{M}_{{\mathrm{P}}{\mathrm{N}};{\mathrm{P}}{\mathrm{P}}} = \mathcal{M}_{{\mathrm{N}}{\mathrm{P}};{\mathrm{N}}{\mathrm{N}}} = \frac{\lambda_m \Xi_\mu^+ \cot(\theta/2)}{\xi_\mu^2}, \\
\mathcal{M}_{{\mathrm{P}}{\mathrm{P}};{\mathrm{N}}{\mathrm{P}}} &=-\mathcal{M}_{{\mathrm{N}}{\mathrm{N}};{\mathrm{P}}{\mathrm{N}}} = \mathcal{M}_{{\mathrm{P}}{\mathrm{N}};{\mathrm{N}}{\mathrm{N}}} = -\mathcal{M}_{{\mathrm{N}}{\mathrm{P}};{\mathrm{P}}{\mathrm{P}}} = -\frac{\Lambda_\mu \cot(\theta/2)}{\xi_\mu^2}, \\
\mathcal{M}_{{\mathrm{P}}{\mathrm{P}};{\mathrm{N}}{\mathrm{N}}} &=\mathcal{M}_{{\mathrm{N}}{\mathrm{N}};{\mathrm{P}}{\mathrm{P}}} = -\mathcal{M}_{{\mathrm{P}}{\mathrm{N}};{\mathrm{N}}{\mathrm{P}}} = -\mathcal{M}_{{\mathrm{N}}{\mathrm{P}};{\mathrm{P}}{\mathrm{N}}} = -\frac{\lambda_m}{\xi_\mu^2}.
\end{align*}

\subsection{Muon-pair production}
\noindent
For $e^-e^+\rightarrow\mu^-\mu^+$, all particles have energy $\omega = \sqrt{|\momentum{p}|^2+m_e^2}$ and the outgoing muons have momentum magnitude $|\momentum{p}_3|=|\momentum{p}_4| = \sqrt{|\momentum{p}|^2+m_e^2-m_\mu^2}$. Defining $\Lambda_\mu = \sqrt{\xi_\mu^2+\lambda_m^2}$, the amplitudes are:
\begin{align*}
\mathcal{M}_{{\mathrm{P}}{\mathrm{P}};{\mathrm{P}}{\mathrm{P}}} &=\mathcal{M}_{{\mathrm{N}}{\mathrm{N}};{\mathrm{N}}{\mathrm{N}}} = -\mathcal{M}_{{\mathrm{P}}{\mathrm{P}};{\mathrm{N}}{\mathrm{N}}} = -\mathcal{M}_{{\mathrm{N}}{\mathrm{N}};{\mathrm{P}}{\mathrm{P}}} = -\frac{\lambda_m \cos\theta}{\Lambda_\mu^2}, \\
\mathcal{M}_{{\mathrm{P}}{\mathrm{P}};{\mathrm{P}}{\mathrm{N}}} &=\mathcal{M}_{{\mathrm{P}}{\mathrm{P}};{\mathrm{N}}{\mathrm{P}}} = -\mathcal{M}_{{\mathrm{N}}{\mathrm{N}};{\mathrm{P}}{\mathrm{N}}} = -\mathcal{M}_{{\mathrm{N}}{\mathrm{N}};{\mathrm{N}}{\mathrm{P}}} = \frac{\lambda_m \sin\theta}{\Lambda_\mu}, \\
\mathcal{M}_{{\mathrm{P}}{\mathrm{N}};{\mathrm{P}}{\mathrm{P}}} &=\mathcal{M}_{{\mathrm{N}}{\mathrm{P}};{\mathrm{P}}{\mathrm{P}}} = -\mathcal{M}_{{\mathrm{P}}{\mathrm{N}};{\mathrm{N}}{\mathrm{N}}} = -\mathcal{M}_{{\mathrm{N}}{\mathrm{P}};{\mathrm{N}}{\mathrm{N}}} = -\frac{\sin\theta}{\Lambda_\mu}, \\
\mathcal{M}_{{\mathrm{P}}{\mathrm{N}};{\mathrm{P}}{\mathrm{N}}} &=\mathcal{M}_{{\mathrm{N}}{\mathrm{P}};{\mathrm{N}}{\mathrm{P}}} = -(1+\cos\theta), \\
\mathcal{M}_{{\mathrm{P}}{\mathrm{N}};{\mathrm{N}}{\mathrm{P}}} &=\mathcal{M}_{{\mathrm{N}}{\mathrm{P}};{\mathrm{P}}{\mathrm{N}}} = 1-\cos\theta.
\end{align*}

To build an outgoing state, we fix $(\xi,\theta)$, read off the amplitudes for each incoming helicity configuration, and insert them into Eq.~\eqref{eq:phi-branches}. The spectator helicity is carried along unchanged.

\begin{figure*}
\centering
\subfloat[Bhabha scattering \label{fig:GGM_bhabha-initial-ghz}]{%
    \includegraphics[width=0.31\linewidth]{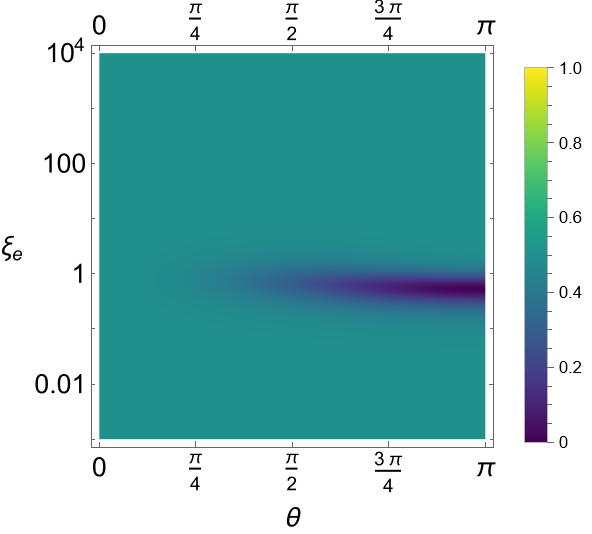}}\hfill
\subfloat[Breit-Wheeler process \label{fig:GGM_breit-wheeler-initial-ghz}]{%
    \includegraphics[width=0.31\linewidth]{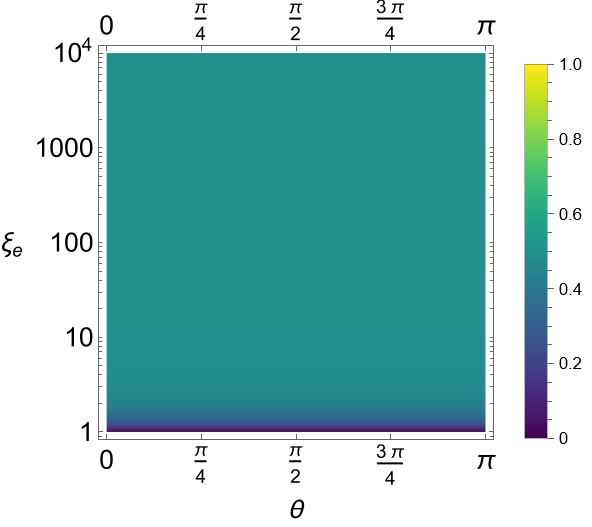}}\hfill
\subfloat[Electron-muon scattering \label{fig:GGM_electron-muon-initial-ghz}]{%
    \includegraphics[width=0.31\linewidth]{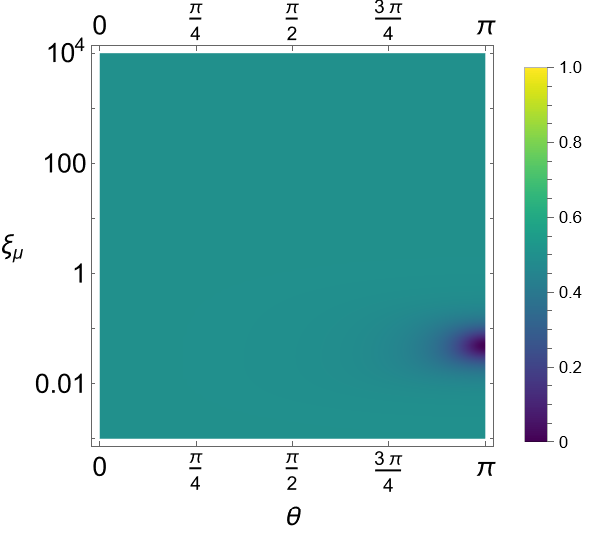}}\\
\subfloat[M\o{}ller scattering \label{fig:GGM_moller-initial-ghz}]{%
    \includegraphics[width=0.31\linewidth]{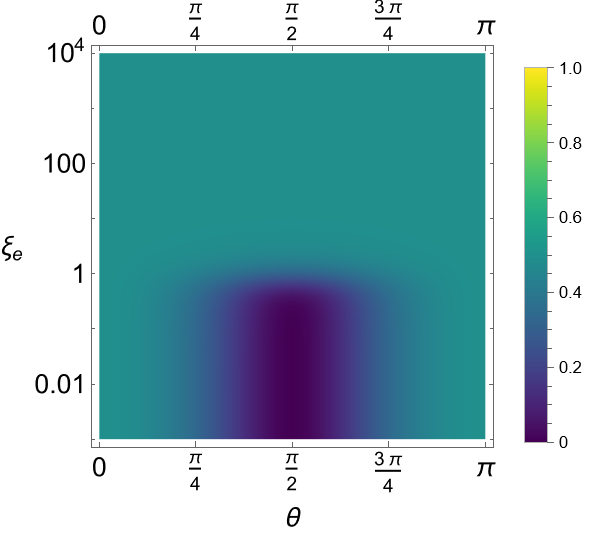}}\hfill
\subfloat[Compton scattering \label{fig:GGM_compton-initial-ghz}]{%
    \includegraphics[width=0.31\linewidth]{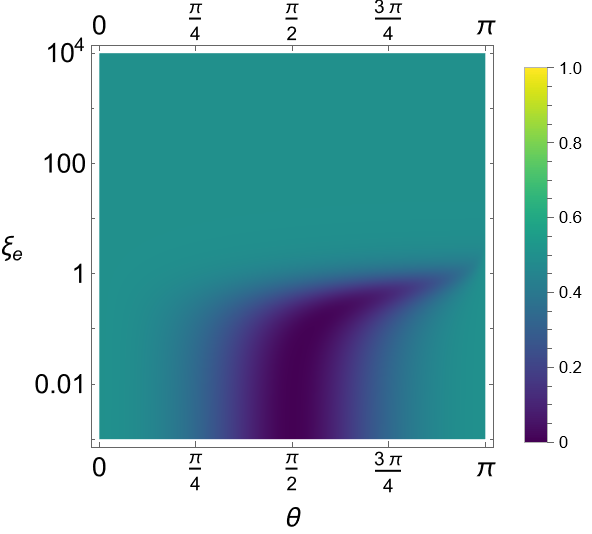}}\hfill
\subfloat[Photon-pair production \label{fig:GGM_photon-pair-production-initial-ghz-app}]{%
    \includegraphics[width=0.31\linewidth]{figures/plots/GGM_Initial-GHZ_photon-pair-production}}
\caption{\label{fig:ghz_ggm} GGM in the $(\theta,\xi)$ plane for an initial GHZ state, arranged as in Fig.~\ref{fig:ghz_tangle}.}
\end{figure*}
\begin{figure*}
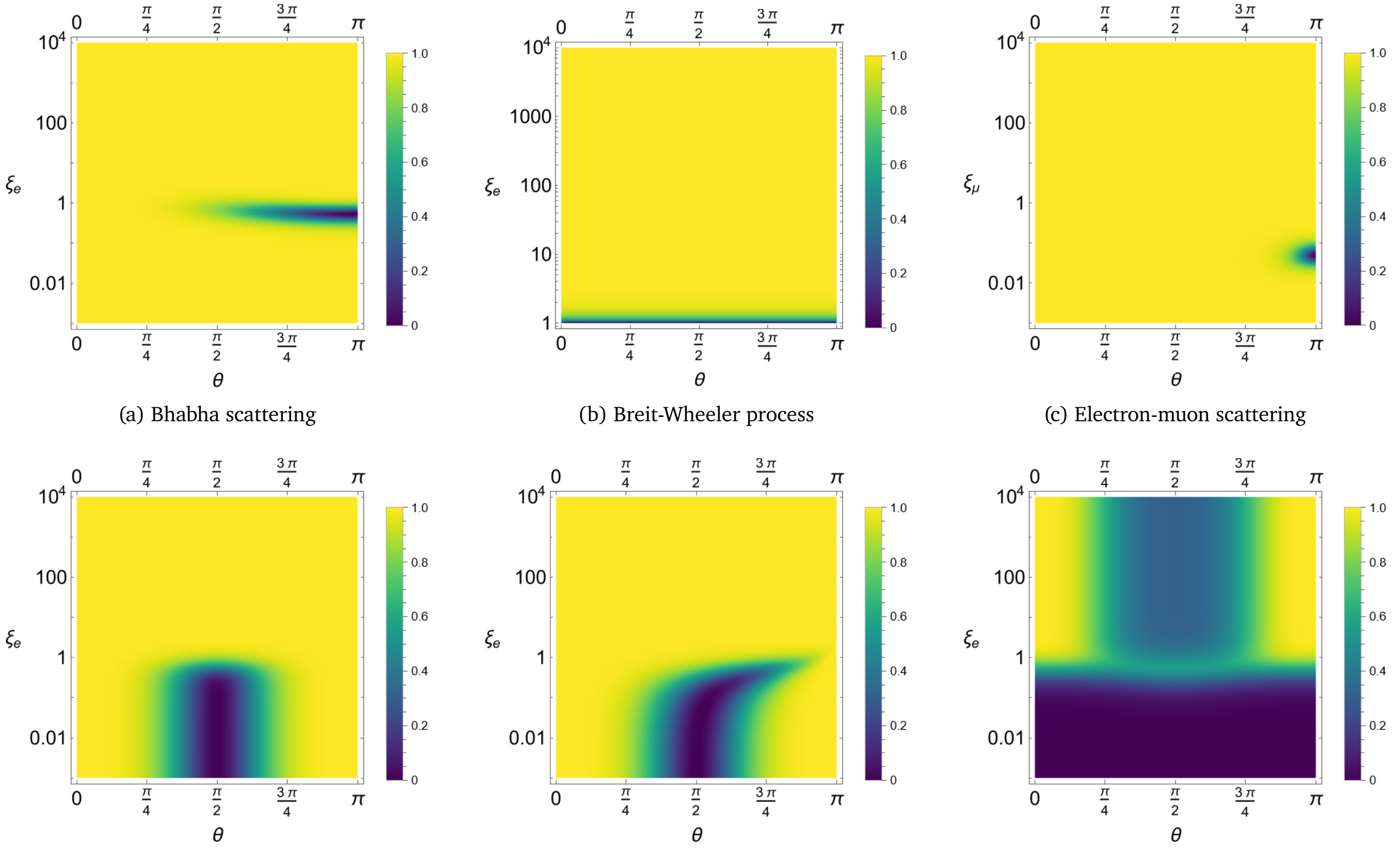

\centering
\subfloat[Bhabha scattering \label{fig:piTangle_bhabha-initial-ghz}]{%
    \includegraphics[width=0.31\linewidth]{figures/plots/pi-Tangle_Initial-GHZ_bhabha}}\hfill
\subfloat[Breit-Wheeler process \label{fig:piTangle_breit-wheeler-initial-ghz}]{%
    \includegraphics[width=0.31\linewidth]{figures/plots/pi-Tangle_Initial-GHZ_breit-wheeler}}\hfill
\subfloat[Electron-muon scattering \label{fig:piTangle_electron-muon-initial-ghz}]{%
    \includegraphics[width=0.31\linewidth]{figures/plots/pi-Tangle_Initial-GHZ_electron-muon}}\\
\subfloat[M\o{}ller scattering \label{fig:piTangle_moller-initial-ghz}]{%
    \includegraphics[width=0.31\linewidth]{figures/plots/pi-Tangle_Initial-GHZ_moller}}\hfill
\subfloat[Compton scattering \label{fig:piTangle_compton-initial-ghz}]{%
    \includegraphics[width=0.31\linewidth]{figures/plots/pi-Tangle_Initial-GHZ_compton}}\hfill
\subfloat[Photon-pair production \label{fig:piTangle_photon-pair-production-initial-ghz-app}]{%
    \includegraphics[width=0.31\linewidth]{figures/plots/pi-Tangle_Initial-GHZ_photon-pair-production}}
\caption{\label{fig:ghz_pitangle} $\pi$-tangle in the $(\theta,\xi)$ plane for an initial GHZ state, arranged as in Fig.~\ref{fig:ghz_tangle}.}
\end{figure*}

\section{Evaluation of GME measures} \label{app:zero-tangle}
\noindent 
We reduce the tangle of the two-branch states in Eq.~\eqref{eq:generic-branches} to a closed expression in terms of the branch amplitudes. We use this to prove the tangle vanishes in Compton scattering and the Breit-Wheeler process for an uncorrelated $A$ input, and to derive the closed forms in Eqs.~\eqref{eq:bw-ghz-tangle} and~\eqref{eq:gg-ghz-tangle} for the GHZ input.

\subsection{Tangle of a two-branch state}
\noindent
We write the outgoing state as
\begin{equation*}
\ket{f}=a\,\ket{\phi_x}_{AB}\ket{{\mathrm{P}}}_C+b\,\ket{\phi_y}_{AB}\ket{{\mathrm{N}}}_C
\end{equation*}
and represent each branch by its $2\times2$ amplitude matrix, $(\Phi_x)_{s_3s_4}=\mathcal{M}_{x;s_3s_4}$, with rows labelled by $s_3$ and columns by $s_4$. For a normalised three-qubit pure state, the tangle is $\tau=4|\operatorname{Hdet}|$~\cite{coffman2000distributed}, where $\operatorname{Hdet}$ is Cayley's hyperdeterminant of the coefficient tensor. Expanding $\operatorname{Hdet}$ shows it equals the discriminant of the quadratic polynomial built from the two slices of the tensor along $C$,
\begin{equation*}
\det\!\left(a\Phi_x+z\,b\Phi_y\right)=a^2\det\Phi_x+z\,ab\,K+z^2\,b^2\det\Phi_y,
\end{equation*}
with the mixed term
\begin{align*}
K&=(\Phi_x)_{{\mathrm{P}}{\mathrm{P}}}(\Phi_y)_{{\mathrm{N}}{\mathrm{N}}}+(\Phi_y)_{{\mathrm{P}}{\mathrm{P}}}(\Phi_x)_{{\mathrm{N}}{\mathrm{N}}}\\
&\quad -(\Phi_x)_{{\mathrm{P}}{\mathrm{N}}}(\Phi_y)_{{\mathrm{N}}{\mathrm{P}}}-(\Phi_y)_{{\mathrm{P}}{\mathrm{N}}}(\Phi_x)_{{\mathrm{N}}{\mathrm{P}}}.
\end{align*}
Including normalisation, we obtain
\begin{align}
\tau_{ABC}&=\frac{4|ab|^2\,|D|}{\left(|a|^2\lVert\Phi_x\rVert^2+|b|^2\lVert\Phi_y\rVert^2\right)^2},\nonumber\\
D&=K^2-4\det\Phi_x\det\Phi_y, \label{eq:tangle-discriminant}
\end{align}
where $\lVert\cdot\rVert$ is the Frobenius norm. We verified Eq.~\eqref{eq:tangle-discriminant} against direct numerical evaluations of the tangle for all seven processes; both agree to machine precision. Because local unitaries multiply $\operatorname{Hdet}$ by a phase, $|D|$ and all subsequent results are independent of the phase conventions in Appendix~\ref{app:qed-helicity-amplitudes}.

For the input with uncorrelated $A$, Eq.~\eqref{eq:fspin}, we have $a=\cos\eta$, $b=\sin\eta$, $\Phi_x=\Phi_{{\mathrm{P}}{\mathrm{P}}}$, and $\Phi_y=\Phi_{{\mathrm{P}}{\mathrm{N}}}$, yielding $4|ab|^2=\sin^2 2\eta$. The discriminant $D$ depends only on the amplitudes, not on $\eta$. The tangle therefore vanishes for every $\eta$ if and only if $D(\xi,\theta)=0$.

\subsection{Breit-Wheeler process with uncorrelated $A$}
\noindent
Recall $\Xi_e^-=\sqrt{\xi_e^2-1}$ and $\Delta=(\xi_e^2-1)\cos^2\theta-\xi_e^2$ from Appendix~\ref{app:qed-helicity-amplitudes}. We have
\begin{align*}
    \Phi_{{\mathrm{P}}{\mathrm{P}}}=\begin{pmatrix}\alpha&0\\0&\beta\end{pmatrix},\qquad
\Phi_{{\mathrm{P}}{\mathrm{N}}}=\begin{pmatrix}g&d\\h&-g\end{pmatrix},   
\end{align*}
with
\begin{align*}
\alpha&=\frac{2(\xi_e+\Xi_e^-)}{\Delta},\qquad \beta=\frac{2(\xi_e-\Xi_e^-)}{\Delta},\qquad g=-\frac{2\Xi_e^-\sin^2\theta}{\Delta},\nonumber\\
d&=\frac{2\xi_e \Xi_e^-\sin\theta(1+\cos\theta)}{\Delta},\qquad h=\frac{2\xi_e \Xi_e^-\sin\theta(\cos\theta-1)}{\Delta}.
\end{align*}
Then $\det\Phi_{{\mathrm{P}}{\mathrm{P}}}=\alpha\beta$, $\det\Phi_{{\mathrm{P}}{\mathrm{N}}}=-g^2-dh$, and $K=g(\beta-\alpha)$, which gives
\begin{align*}
D&=g^2(\beta-\alpha)^2+4\alpha\beta\,(g^2+dh)\\
&=g^2(\alpha+\beta)^2+4\alpha\beta\,dh.
\end{align*}
With $\alpha+\beta=4\xi_e/\Delta$, $\alpha\beta=4(\xi_e^2-(\Xi_e^-)^2)/\Delta^2=4/\Delta^2$, and $dh=-4\xi_e^2(\Xi_e^-)^2\sin^4\theta/\Delta^2$, the two terms become
\begin{align*}
g^2(\alpha+\beta)^2&=\frac{64\,\xi_e^2(\Xi_e^-)^2\sin^4\theta}{\Delta^4},\\
4\alpha\beta\,dh&=-\frac{64\,\xi_e^2(\Xi_e^-)^2\sin^4\theta}{\Delta^4},
\end{align*}
which cancel exactly. Hence $D\equiv0$, and $\tau_{ABC}=0$ for all $\xi_e\geq1$, $\theta$, and $\eta$.

\subsection{Compton scattering with uncorrelated $A$}
\noindent
Let $c=\cos(\theta/2)$, $s=\sin(\theta/2)$, $\Xi_e^+=\sqrt{\xi_e^2+1}$, $w=\xi_e+\Xi_e^+$, such that $\Xi_e^+-\xi_e=1/w$, and $E=\xi_e\cos\theta+\Xi_e^+$. Three identities simplify the amplitudes: $4\xi_e w=2(w^2-1)$, $1+\cos\theta=2c^2$, and $(\Xi_e^+-\xi_e\cos\theta)E=1+\xi_e^2\sin^2\theta$. Every amplitude then acquires a common factor of $2/E$, which drops out of the condition $D=0$. Without it, the branch matrices are
\begin{align*}
\Phi_{{\mathrm{P}}{\mathrm{P}}}\propto\begin{pmatrix}-\dfrac{c(w^2-s^2)}{w}&-sc^2\\[2mm]-\dfrac{s^2c}{w}&-\dfrac{s^3}{w^2}\end{pmatrix},\qquad
\Phi_{{\mathrm{P}}{\mathrm{N}}}\propto\begin{pmatrix}sc^2&-wc^3\\[1mm]s^3&-\dfrac{s^2c}{w}\end{pmatrix}.
\end{align*}
The three coefficients of the quadratic are
\begin{align*}
\det\Phi_{{\mathrm{P}}{\mathrm{P}}}&\propto\frac{s^3c}{w^3}\left[(w^2-s^2)-c^2w^2\right]=\frac{s^5c\,(w^2-1)}{w^3},\nonumber\\
\det\Phi_{{\mathrm{P}}{\mathrm{N}}}&\propto-\frac{s^3c^3}{w}+w\,s^3c^3=\frac{s^3c^3(w^2-1)}{w},\nonumber\\
K&\propto s^2c^2\left[\frac{w^2-s^2}{w^2}-\frac{s^2}{w^2}+s^2-c^2\right]=\frac{2s^4c^2(w^2-1)}{w^2},
\end{align*}
where the final step uses $1-c^2+s^2=2s^2$. Therefore,
\begin{align*}
K^2=\frac{4s^8c^4(w^2-1)^2}{w^4}=4\det\Phi_{{\mathrm{P}}{\mathrm{P}}}\det\Phi_{{\mathrm{P}}{\mathrm{N}}},
\end{align*}
so again $D\equiv0$, and $\tau_{ABC}=0$ for all $\xi_e$, $\theta$, and $\eta$.

Since a pure state with zero tangle belongs to the W class when it carries GME~\cite{dur2000three}, and both the GGM and $\pi$-tangle confirm the presence of GME [Figs.~\ref{fig:GGM_breit-wheeler}, \ref{fig:GGM_compton}, \ref{fig:pi-Tangle_breit-wheeler}, and~\ref{fig:pi-Tangle_compton}], both processes produce exclusively W-class entanglement.

\subsection{Breit-Wheeler process with the GHZ input}
\noindent
For the GHZ input, Eq.~\eqref{eq:fghz}, $a=b=1/\sqrt{2}$. Writing $a_+=\mathcal{M}_{{\mathrm{P}}{\mathrm{P}};{\mathrm{P}}{\mathrm{P}}}$ and $a_-=\mathcal{M}_{{\mathrm{P}}{\mathrm{P}};{\mathrm{N}}{\mathrm{N}}}$, the branch matrices are $\Phi_{{\mathrm{P}}{\mathrm{P}}}=\mathrm{diag}(a_+,a_-)$ and $\Phi_{{\mathrm{N}}{\mathrm{N}}}=-\mathrm{diag}(a_-,a_+)$, leading to
\begin{align*}
\det\!\left(\Phi_{{\mathrm{P}}{\mathrm{P}}}+z\,\Phi_{{\mathrm{N}}{\mathrm{N}}}\right)&=a_+a_--z\left(a_+^2+a_-^2\right)+z^2a_+a_-,\\
\qquad D&=\left(a_+^2-a_-^2\right)^2.
\end{align*}
With $\lVert\Phi_{{\mathrm{P}}{\mathrm{P}}}\rVert^2=\lVert\Phi_{{\mathrm{N}}{\mathrm{N}}}\rVert^2=a_+^2+a_-^2$, Eq.~\eqref{eq:tangle-discriminant} gives
\begin{align*}
\tau_{ABC}=\left(\frac{a_+^2-a_-^2}{a_+^2+a_-^2}\right)^2=\left(\frac{2\xi_e\Xi_e^-}{2\xi_e^2-1}\right)^2,
\end{align*}
since $a_\pm\propto\xi_e\pm\Xi_e^-$ with a common $\theta$-dependent factor that cancels. This recovers Eq.~\eqref{eq:bw-ghz-tangle}. It is independent of $\theta$ and saturates the bound in Eq.~\eqref{eq:ghz-bound} since $\braket{\phi_{\mathrm{PP}}|\phi_{\mathrm{NN}}}=-2a_+a_-$.

\subsection{Photon-pair production with the GHZ input}
\noindent
Again $a=b=1/\sqrt{2}$. With $\Xi_e^+=\sqrt{\xi_e^2+1}$, $c=\xi_e\sin^2\theta$, and $\Delta_\gamma=\xi_e^2\sin^2\theta+1$, the amplitudes yield
\begin{align*}
&\Phi_{{\mathrm{P}}{\mathrm{P}}}=\frac{2}{\Delta_\gamma}\begin{pmatrix}-(\xi_e+\Xi_e^+) & c\\ c & \Xi_e^+-\xi_e\end{pmatrix},\\&
\Phi_{{\mathrm{N}}{\mathrm{N}}}=\frac{2}{\Delta_\gamma}\begin{pmatrix}-(\Xi_e^+-\xi_e) & -c\\ -c & \xi_e+\Xi_e^+\end{pmatrix}.
\end{align*}
The common factor $2/\Delta_\gamma$ cancels in Eq.~\eqref{eq:tangle-discriminant}. Using $(\xi_e+\Xi_e^+)(\Xi_e^+-\xi_e)=1$, we find
\begin{align*}
&\det\Phi_{{\mathrm{P}}{\mathrm{P}}}=\det\Phi_{{\mathrm{N}}{\mathrm{N}}}=-(1+c^2),\\
& K=-(\xi_e+\Xi_e^+)^2-(\Xi_e^+-\xi_e)^2+2c^2=-2\left(2\xi_e^2+1-c^2\right),
\end{align*}
and therefore
\begin{align*}
D&=4\left(2\xi_e^2+1-c^2\right)^2-4\left(1+c^2\right)^2\\&=16\left(\xi_e^2-c^2\right)\left(\xi_e^2+1\right)\\&=16\,\xi_e^2\left(\xi_e^2+1\right)\cos^2\theta\left(1+\sin^2\theta\right).
\end{align*}
With $\lVert\Phi_{{\mathrm{P}}{\mathrm{P}}}\rVert^2=\lVert\Phi_{{\mathrm{N}}{\mathrm{N}}}\rVert^2=2\left(2\xi_e^2+1+c^2\right)$, Eq.~\eqref{eq:tangle-discriminant} produces Eq.~\eqref{eq:gg-ghz-tangle}. The tangle vanishes at $\theta=\pi/2$ for every $\xi_e$, although the branches are not parallel there. Since $\braket{\phi_{\mathrm{PP}}|\phi_{\mathrm{NN}}}\propto 2(1-c^2)$, Eq.~\eqref{eq:ghz-bound} only requires $\tau_{ABC}\leq1-\left[(1-\xi_e^2)/(3\xi_e^2+1)\right]^2$ at $\theta=\pi/2$, which approaches $8/9$ at high momentum.

At $\theta=\pi/2$ and $\xi_e=1/\sqrt{3}$, the normalised outgoing state is, with qubit order $ABC$,
\begin{align*}
\ket{f}=\frac{1}{\sqrt{24}}&\big(-3\ket{{\mathrm{P}}{\mathrm{P}}{\mathrm{P}}}-\ket{{\mathrm{P}}{\mathrm{P}}{\mathrm{N}}}+\ket{{\mathrm{P}}{\mathrm{N}}{\mathrm{P}}}-\ket{{\mathrm{P}}{\mathrm{N}}{\mathrm{N}}}\\&+\ket{{\mathrm{N}}{\mathrm{P}}{\mathrm{P}}}-\ket{{\mathrm{N}}{\mathrm{P}}{\mathrm{N}}}+\ket{{\mathrm{N}}{\mathrm{N}}{\mathrm{P}}}+3\ket{{\mathrm{N}}{\mathrm{N}}{\mathrm{N}}}\big).
\end{align*}
Mapping $\ket{{\mathrm{P}}} \to \ket{0}$ and $\ket{{\mathrm{N}}} \to \ket{1}$, a direct calculation yields $(XH\otimes XH\otimes H)\ket{f}=-(\ket{001}+\ket{010}+\ket{100})/\sqrt{3}$, where $H$ is the Hadamard gate and $X$ the Pauli $X$ gate. Thus, the outgoing state is locally unitarily equivalent to the W state, carrying $\tau_{ABC}=0$, $\mathcal{G}_{ABC}=1/3$, and $\pi_{ABC}\approx 0.55$.

Figures~\ref{fig:ghz_ggm} and~\ref{fig:ghz_pitangle} complete Fig.~\ref{fig:ghz_tangle}. Except in photon-pair production, the $\pi$-tangle coincides with the tangle to plotting accuracy, with the GGM tracking the identical pattern over its $[0,1/2]$ range.
\end{document}

%% file: figures/qed_scattering.tex
\begin{tikzpicture}[
    scale=0.68, 
    thick,
    >=Latex,
    fermion/.style={NavyBlue!70!black, line width=1.2pt},
    ent/.style={dashed, black!70},
]

\begin{scope}[shift={(0,0)}]
    \draw[fermion, ->] (-2,-3) -- (-0.8,-0.8);
    \draw[fermion, ->] (-0.8,0.8) -- (-2,3);
    
    \draw[fermion, ->] (0.8,-3) -- (0.2,-0.8);
    \draw[fermion, ->] (0.2,0.8) -- (0.8,3);
    
    \draw[fermion, ->] (2,-3) -- (2,3);
    
    \draw[fill=olive!50!white] (-0.2,0) ellipse (1.9 and 0.7);
    \node at (-0.2,0) {Scattering};
    
    \node at (-2.3,3.5) {$A'$};
    \node at (-2.3,-3.5) {$A$};
    \node at (1.0,3.5) {$B'$};
    \node at (1.0,-3.5) {$B$};
    \node at (2,3.5) {$C$};
    \node at (2,-3.5) {$C$};
    
    \draw[ent] (0.2,1.8) ellipse (2.5 and 0.5);
    
    \draw[ent]
        (1.3,-2)
        .. controls (1.8,-1.6) and (2.5,-1.6)
        .. (2.5,-2.2)
        .. controls (2.5,-2.8) and (1.8,-2.8)
        .. (1.3,-2.2)
        .. controls (0.8,-1.6) and (0.1,-1.6)
        .. (0.1,-2.2)
        .. controls (0.1,-2.8) and (0.8,-2.8)
        .. (1.3,-2.2);
\end{scope}

\begin{scope}[shift={(7,0)}]
    \draw[fermion, ->] (-2,-3) -- (-0.8,-0.8);
    \draw[fermion, ->] (-0.8,0.8) -- (-2,3);
    
    \draw[fermion, ->] (0.8,-3) -- (0.2,-0.8);
    \draw[fermion, ->] (0.2,0.8) -- (0.8,3);
    
    \draw[fermion, ->] (2,-3) -- (2,3);
    
    \draw[fill=olive!50!white] (-0.2,0) ellipse (1.9 and 0.7);
    \node at (-0.2,0) {Scattering};
    
    \node at (-2.3,3.5) {$A'$};
    \node at (-2.3,-3.5) {$A$};
    \node at (1.0,3.5) {$B'$};
    \node at (1.0,-3.5) {$B$};
    \node at (2,3.5) {$C$};
    \node at (2,-3.5) {$C$};
    
    \draw[ent] (0.2,1.8) ellipse (2.5 and 0.5);
    
    \draw[ent] (0.2,-2.2) ellipse (2.5 and 0.5);
\end{scope}

\end{tikzpicture}

%% file: apssamp.bib
@article{abrol2025maximal,
  title = {Maximal Secret Reconstruction, Teleportation and {{Bell}}'s Inequality},
  author = {Abrol, Pratishtha and Singh, Pahulpreet and Chakrabarty, Indranil},
  year = 2025,
  month = feb,
  journal = {The European Physical Journal D},
  volume = {79},
  number = {2},
  pages = {11},
  issn = {1434-6060, 1434-6079},
  doi = {10.1140/epjd/s10053-025-00955-6},
  langid = {english}
}

@article{araujo2019measuring,
  title = {Measuring {{QED}} Cross Sections via Entanglement},
  author = {Araujo, Jonas B. and Hiller, B. and Da Paz, I. G. and Ferreira, Manoel M. and Sampaio, Marcos and Costa, H. A. S.},
  year = 2019,
  month = nov,
  journal = {Physical Review D},
  volume = {100},
  number = {10},
  pages = {105018},
  issn = {2470-0010, 2470-0029},
  doi = {10.1103/PhysRevD.100.105018},
  langid = {english}
}

@article{bennett1996concentrating,
  title = {Concentrating Partial Entanglement by Local Operations},
  author = {Bennett, Charles H. and Bernstein, Herbert J. and Popescu, Sandu and Schumacher, Benjamin},
  year = 1996,
  month = apr,
  journal = {Physical Review A},
  volume = {53},
  number = {4},
  pages = {2046--2052},
  issn = {1050-2947, 1094-1622},
  doi = {10.1103/PhysRevA.53.2046},
  langid = {english}
}

@article{blasone2024entanglement,
  title = {Entanglement Distribution in {{Bhabha}} Scattering with an Entangled Spectator Particle},
  author = {Blasone, Massimo and Lambiase, Gaetano and Micciola, Bruno},
  year = 2024,
  month = may,
  journal = {Physical Review D},
  volume = {109},
  number = {9},
  pages = {096022},
  issn = {2470-0010, 2470-0029},
  doi = {10.1103/PhysRevD.109.096022},
  langid = {english}
}

@article{blasone2025complete,
  title = {Complete Complementarity Relations in Tree Level {{QED}} Processes},
  author = {Blasone, Massimo and De Siena, Silvio and Lambiase, Gaetano and Matrella, Cristina and Micciola, Bruno},
  year = 2025,
  month = jan,
  journal = {Physical Review D},
  volume = {111},
  number = {1},
  pages = {016007},
  issn = {2470-0010, 2470-0029},
  doi = {10.1103/PhysRevD.111.016007},
  langid = {english}
}

@article{blasone2025entanglement,
  title = {Entanglement Dynamics in {{QED}} Processes},
  author = {Blasone, M. and De Siena, S. and Lambiase, G. and Matrella, C. and Micciola, B.},
  year = 2025,
  month = jun,
  journal = {Chaos, Solitons \& Fractals},
  volume = {195},
  pages = {116305},
  issn = {09600779},
  doi = {10.1016/j.chaos.2025.116305},
  langid = {english}
}

@article{cao2026genuine,
  title = {Genuine Tripartite Entanglement in {{Bhabha}} Scattering with an Entangled Spectator Particle},
  author = {Cao, Zan and Song, Meng-Long and Song, Xue-Ke and Ye, Liu and Wang, Dong},
  year = 2026,
  month = may,
  journal = {Physical Review D},
  volume = {113},
  number = {10},
  pages = {105019},
  issn = {2470-0010, 2470-0029},
  doi = {10.1103/1885-yhgf},
  langid = {english}
}

@article{carney2016scattering,
    author = "Carney, Daniel and Chaurette, Laurent and Semenoff, Gordon",
    title = "{Scattering with partial information}",
    eprint = "1606.03103",
    archivePrefix = "arXiv",
    primaryClass = "hep-th",
    month = "6",
    year = "2016"
}

@article{cervera-lierta2017maximal,
  title = {Maximal {{Entanglement}} in {{High Energy Physics}}},
  author = {{Cervera-Lierta}, Alba and Latorre, Jos{\'e} and Rojo, Juan and Rottoli, Luca},
  year = 2017,
  month = nov,
  journal = {SciPost Physics},
  volume = {3},
  number = {5},
  pages = {036},
  issn = {2542-4653},
  doi = {10.21468/SciPostPhys.3.5.036}
}

@article{coffman2000distributed,
  title = {Distributed Entanglement},
  author = {Coffman, Valerie and Kundu, Joydip and Wootters, William K.},
  year = 2000,
  month = apr,
  journal = {Physical Review A},
  volume = {61},
  number = {5},
  pages = {052306},
  issn = {1050-2947, 1094-1622},
  doi = {10.1103/PhysRevA.61.052306},
  langid = {english}
}

@article{dur2000three,
  title = {Three Qubits Can Be Entangled in Two Inequivalent Ways},
  author = {D{\"u}r, W. and Vidal, G. and Cirac, J. I.},
  year = 2000,
  month = nov,
  journal = {Physical Review A},
  volume = {62},
  number = {6},
  pages = {062314},
  issn = {1050-2947, 1094-1622},
  doi = {10.1103/PhysRevA.62.062314},
  langid = {english}
}

@article{faleiro2020perturbative,
  title = {Perturbative Approach to Entanglement Generation in {{QFT}} Using the {{S}} Matrix},
  author = {Faleiro, Ricardo and Pav{\~a}o, Rafael and Costa, Helder A S and Hiller, Brigitte and Blin, Alex H and Sampaio, Marcos},
  year = 2020,
  month = sep,
  journal = {Journal of Physics A: Mathematical and Theoretical},
  volume = {53},
  number = {36},
  pages = {365301},
  issn = {1751-8113, 1751-8121},
  doi = {10.1088/1751-8121/aba214}
}

@article{fan2017variation,
  title = {Variation of Entanglement Entropy and Mutual Information in Fermion-Fermion Scattering},
  author = {Fan, Jinbo and Deng, Yanbin and Huang, Yong-Chang},
  year = 2017,
  month = mar,
  journal = {Physical Review D},
  volume = {95},
  number = {6},
  pages = {065017},
  issn = {2470-0010, 2470-0029},
  doi = {10.1103/PhysRevD.95.065017},
  langid = {english}
}

@article{fedida2023treelevel,
  title = {Tree-Level Entanglement in Quantum Electrodynamics},
  author = {Fedida, Samuel and Serafini, Alessio},
  year = 2023,
  month = jun,
  journal = {Physical Review D},
  volume = {107},
  number = {11},
  pages = {116007},
  issn = {2470-0010, 2470-0029},
  doi = {10.1103/PhysRevD.107.116007},
  langid = {english}
}

@article{fonseca2022entanglement,
  title = {Entanglement and Scattering in Quantum Electrodynamics: {{S}} Matrix Information from an Entangled Spectator Particle},
  shorttitle = {Entanglement and Scattering in Quantum Electrodynamics},
  author = {Fonseca, Juan D. and Hiller, B. and Araujo, J. B. and Da Paz, I. G. and Sampaio, M.},
  year = 2022,
  month = sep,
  journal = {Physical Review D},
  volume = {106},
  number = {5},
  pages = {056015},
  issn = {2470-0010, 2470-0029},
  doi = {10.1103/PhysRevD.106.056015},
  langid = {english}
}

@article{guhne2009entanglement,
  title = {Entanglement Detection},
  author = {G{\"u}hne, Otfried and T{\'o}th, G{\'e}za},
  year = 2009,
  month = apr,
  journal = {Physics Reports},
  volume = {474},
  number = {1-6},
  pages = {1--75},
  issn = {03701573},
  doi = {10.1016/j.physrep.2009.02.004},
  langid = {english}
}

@article{hamzehofi2026entanglement,
  title = {Entanglement Measures for Multipartite Quantum Systems: Limitations and New Approaches},
  shorttitle = {Entanglement Measures for Multipartite Quantum Systems},
  author = {Hamzehofi, Reza},
  year = 2026,
  month = may,
  journal = {The European Physical Journal D},
  volume = {80},
  number = {5},
  pages = {67},
  issn = {1434-6060, 1434-6079},
  doi = {10.1140/epjd/s10053-026-01182-3},
  langid = {english}
}

@article{jacob1959general,
  title = {On the General Theory of Collisions for Particles with Spin},
  author = {Jacob, M. and Wick, G.C.},
  year = 1959,
  month = aug,
  journal = {Annals of Physics},
  volume = {7},
  number = {4},
  pages = {404--428},
  issn = {00034916},
  doi = {10.1016/0003-4916(59)90051-X},
  langid = {english}
}

@article{li2017tight,
  title = {Tight Upper Bound for the Maximal Quantum Value of the {{Svetlichny}} Operators},
  author = {Li, Ming and Shen, Shuqian and Jing, Naihuan and Fei, Shao-Ming and {Li-Jost}, Xianqing},
  year = 2017,
  month = oct,
  journal = {Physical Review A},
  volume = {96},
  number = {4},
  pages = {042323},
  issn = {2469-9926, 2469-9934},
  doi = {10.1103/PhysRevA.96.042323},
  langid = {english}
}

@article{liang2007bounds,
  title = {Bounds on Quantum Correlations in {{Bell}}-Inequality Experiments},
  author = {Liang, Yeong-Cherng and Doherty, Andrew C.},
  year = 2007,
  month = apr,
  journal = {Physical Review A},
  volume = {75},
  number = {4},
  pages = {042103},
  issn = {1050-2947, 1094-1622},
  doi = {10.1103/PhysRevA.75.042103},
  langid = {english}
}

@article{manoukian2004speed,
  title = {Speed Dependent Polarization Correlations in {{QED}} and Entanglement},
  author = {Manoukian, E. B. and Yongram, N.},
  year = 2004,
  month = oct,
  journal = {The European Physical Journal D},
  volume = {31},
  number = {1},
  pages = {137--143},
  issn = {1434-6060, 1434-6079},
  doi = {10.1140/epjd/e2004-00131-2},
  langid = {english}
}

@misc{muhuri2026characterizing,
    author = "Muhuri, Abhishek and Ganguly, Nirman and De, Aditi Sen and Chakrabarty, Indranil",
    title = "{Characterizing pairwise swapping capabilities of dense coding channels}",
    eprint = "2608.03184",
    archivePrefix = "arXiv",
    primaryClass = "quant-ph",
    month = "8",
    year = "2026"
}

@article{ou2007monogamy,
  title = {Monogamy Inequality in Terms of Negativity for Three-Qubit States},
  author = {Ou, Yong-Cheng and Fan, Heng},
  year = 2007,
  month = jun,
  journal = {Physical Review A},
  volume = {75},
  number = {6},
  pages = {062308},
  issn = {1050-2947, 1094-1622},
  doi = {10.1103/PhysRevA.75.062308},
  langid = {english}
}

@article{pandya2016complementarity,
  title = {Complementarity between Tripartite Quantum Correlation and Bipartite {{Bell}}-Inequality Violation in Three-Qubit States},
  author = {Pandya, Palash and Misra, Avijit and Chakrabarty, Indranil},
  year = 2016,
  month = nov,
  journal = {Physical Review A},
  volume = {94},
  number = {5},
  pages = {052126},
  issn = {2469-9926, 2469-9934},
  doi = {10.1103/PhysRevA.94.052126},
  langid = {english}
}

@article{peschanski2016entanglement,
  title = {Entanglement Entropy of Scattering Particles},
  author = {Peschanski, Robi and Seki, Shigenori},
  year = 2016,
  month = jul,
  journal = {Physics Letters B},
  volume = {758},
  pages = {89--92},
  issn = {03702693},
  doi = {10.1016/j.physletb.2016.04.063},
  langid = {english}
}

@article{peschanski2019evaluation,
  title = {Evaluation of Entanglement Entropy in High Energy Elastic Scattering},
  author = {Peschanski, Robi and Seki, Shigenori},
  year = 2019,
  month = oct,
  journal = {Physical Review D},
  volume = {100},
  number = {7},
  pages = {076012},
  issn = {2470-0010, 2470-0029},
  doi = {10.1103/PhysRevD.100.076012},
  langid = {english}
}

@article{prabhu2013exclusion,
  title = {Exclusion Principle for Quantum Dense Coding},
  author = {Prabhu, R. and Pati, Arun Kumar and Sen(De), Aditi and Sen, Ujjwal},
  year = 2013,
  month = may,
  journal = {Physical Review A},
  volume = {87},
  number = {5},
  pages = {052319},
  issn = {1050-2947, 1094-1622},
  doi = {10.1103/PhysRevA.87.052319},
  langid = {english}
}

@article{quinta2024multipartite,
  title = {Multipartite Entanglement from Consecutive Scatterings},
  author = {Quinta, Gon{\c c}alo M. and Andr{\'e}, Rui},
  year = 2024,
  month = feb,
  journal = {Physical Review A},
  volume = {109},
  number = {2},
  pages = {022433},
  issn = {2469-9926, 2469-9934},
  doi = {10.1103/PhysRevA.109.022433},
  langid = {english}
}

@article{sakurai2024threebody,
  title = {Three-{{Body Entanglement}} in {{Particle Decays}}},
  author = {Sakurai, Kazuki and Spannowsky, Michael},
  year = 2024,
  month = apr,
  journal = {Physical Review Letters},
  volume = {132},
  number = {15},
  pages = {151602},
  issn = {0031-9007, 1079-7114},
  doi = {10.1103/PhysRevLett.132.151602},
  langid = {english}
}

@article{sazim2015retrieving,
  title = {Retrieving and Routing Quantum Information in a Quantum Network},
  author = {Sazim, S. and Chiranjeevi, V. and Chakrabarty, I. and Srinathan, K.},
  year = 2015,
  month = dec,
  journal = {Quantum Information Processing},
  volume = {14},
  number = {12},
  pages = {4651--4664},
  issn = {1570-0755, 1573-1332},
  doi = {10.1007/s11128-015-1109-7},
  langid = {english}
}

@article{seki2015variation,
  title = {Variation of Entanglement Entropy in Scattering Process},
  author = {Seki, Shigenori and Park, I.Y. and Sin, Sang-Jin},
  year = 2015,
  month = apr,
  journal = {Physics Letters B},
  volume = {743},
  pages = {147--153},
  issn = {03702693},
  doi = {10.1016/j.physletb.2015.02.028},
  langid = {english}
}

@article{sende2010channel,
  title = {Channel Capacities versus Entanglement Measures in Multiparty Quantum States},
  author = {Sen(De), Aditi and Sen, Ujjwal},
  year = 2010,
  month = jan,
  journal = {Physical Review A},
  volume = {81},
  number = {1},
  pages = {012308},
  issn = {1050-2947, 1094-1622},
  doi = {10.1103/PhysRevA.81.012308},
  langid = {english}
}

@article{singh2024controlled,
  title = {Controlled State Reconstruction and Quantum Secret Sharing},
  author = {Singh, Pahulpreet and Chakrabarty, Indranil},
  year = 2024,
  month = mar,
  journal = {Physical Review A},
  volume = {109},
  number = {3},
  pages = {032406},
  issn = {2469-9926, 2469-9934},
  doi = {10.1103/PhysRevA.109.032406},
  langid = {english}
}

@article{srivastava2019oneshot,
  title = {One-Shot Conclusive Multiport Quantum Dense Coding Capacities},
  author = {Srivastava, Chirag and Bera, Anindita and Sen(De), Aditi and Sen, Ujjwal},
  year = 2019,
  month = nov,
  journal = {Physical Review A},
  volume = {100},
  number = {5},
  pages = {052304},
  issn = {2469-9926, 2469-9934},
  doi = {10.1103/PhysRevA.100.052304},
  langid = {english}
}

@article{svetlichny1987distinguishing,
  title = {Distinguishing Three-Body from Two-Body Nonseparability by a {{Bell}}-Type Inequality},
  author = {Svetlichny, George},
  year = 1987,
  month = may,
  journal = {Physical Review D},
  volume = {35},
  number = {10},
  pages = {3066--3069},
  issn = {0556-2821},
  doi = {10.1103/PhysRevD.35.3066},
  langid = {english}
}

@article{werner2001bell,
    author = "Werner, Reinhard F. and Wolf, Michael M.",
    title = "{Bell inequalities and entanglement}",
    eprint = "quant-ph/0107093",
    archivePrefix = "arXiv",
    doi = "10.26421/QIC1.3-1",
    journal = "Quant. Inf. Comput.",
    volume = "1",
    number = "3",
    pages = "1--25",
    year = "2001"
}

@article{yang2004efficient,
  title = {Efficient Many-Party Controlled Teleportation of Multiqubit Quantum Information via Entanglement},
  author = {Yang, Chui-Ping and Chu, Shih-I and Han, Siyuan},
  year = 2004,
  month = aug,
  journal = {Physical Review A},
  volume = {70},
  number = {2},
  pages = {022329},
  issn = {1050-2947, 1094-1622},
  doi = {10.1103/PhysRevA.70.022329},
  langid = {english}
}

@article{yongram2008spin,
  title = {Spin Correlations in Elastic $e^+e^-$ Scattering in {{QED}}},
  author = {Yongram, N.},
  year = 2008,
  month = apr,
  journal = {The European Physical Journal D},
  volume = {47},
  number = {1},
  pages = {71--74},
  issn = {1434-6060, 1434-6079},
  doi = {10.1140/epjd/e2008-00030-6},
  langid = {english}
}

@article{yongram2011spin,
  title = {Spin {{Correlations}} in $e^+ e^-$ {{Pair Creation}} by {{Two-Photons}} and {{Entanglement}} in {{QED}}},
  author = {Yongram, N.},
  year = 2011,
  month = mar,
  journal = {International Journal of Theoretical Physics},
  volume = {50},
  number = {3},
  pages = {838--849},
  issn = {0020-7748, 1572-9575},
  doi = {10.1007/s10773-010-0621-x},
  langid = {english}
}
